%% file: Zaid0324.tex
\documentclass[twocolumn,twocolappendix]{aastex631}

\usepackage{xcolor}
\usepackage{amsmath}
\usepackage{multirow}
\usepackage{graphicx}
\usepackage{xspace}
\usepackage{subfigure}
\usepackage{changepage}

\newcommand{\VAST}{\texttt{VAST}\xspace}
\newcommand{\VV}{\texttt{V$^2$}\xspace}
\newcommand{\VF}{\texttt{VoidFinder}\xspace}

\newcommand{\hMpc}{$h^{-1}$~Mpc\xspace}
\newcommand{\cmmnt}[1]{}

\begin{document}

\title{The impact of void-finding algorithms on galaxy classification}

\shorttitle{Differences in void galaxy classification}
\shortauthors{Zaidouni et al.}

\author[0000-0003-0931-0868]{Fatima Zaidouni}
\altaffiliation{Now at Department of Physics, Massachusetts Institute of Technology, Cambridge, MA 02139}
\affiliation{Department of Physics and Astronomy, University of Rochester, Rochester, NY USA 14627}

\author[0000-0001-8101-2836]{Dahlia Veyrat}
\affiliation{Department of Physics and Astronomy, University of Rochester, Rochester, NY USA 14627}

\author[0000-0002-9540-546X]{Kelly A. Douglass}
\affiliation{Department of Physics and Astronomy, University of Rochester, Rochester, NY USA 14627}
\correspondingauthor{Kelly A. Douglass}
\email{kellyadouglass@rochester.edu}

\author[0000-0001-5537-4710]{Segev BenZvi}
\affiliation{Department of Physics and Astronomy, University of Rochester, Rochester, NY USA 14627}

\begin{abstract}
  We explore how the definition of a void influences the conclusions drawn about 
  the impact of the void environment on galactic properties using two 
  void-finding algorithms in the Void Analysis Software Toolkit: \VV, a Python 
  implementation of ZOBOV, and \VF, an algorithm which grows and merges 
  spherical void regions.  Using the Sloan Digital Sky Survey Data Release 7, we 
  find that galaxies found in \VF voids tend to be bluer, fainter, and have 
  higher (specific) star formation rates than galaxies in denser regions.  
  Conversely, galaxies found in \VV voids show no significant differences when 
  compared to galaxies in denser regions, inconsistent with the large-scale 
  environmental effects on galaxy properties expected from both simulations and 
  previous observations.  These results align with previous simulation results 
  that show \VV-identified voids ``leaking'' into the dense walls between voids 
  because their boundaries extend up to the density maxima in the walls.  As a 
  result, when using ZOBOV-based void finders, galaxies likely to be part of 
  wall regions are instead classified as void galaxies, a misclassification that 
  can be critical to our understanding of galaxy evolution.
\end{abstract}

\section{Introduction}\label{sec:intro}

The large-scale structure of the universe observed in galaxy redshift surveys 
exhibits a cosmic web-like structure \citep{Bond96} comprising voids, or vast 
underdense regions tens of megaparsecs across, that are bordered by filaments of 
galaxies that flow into large galaxy clusters \citep{deLapparent86}.  Because of 
the low matter density within voids, their gravitational evolution remains in 
the linear regime \citep{Sutter12a}, and the few galaxies found in voids evolve 
relatively free of interactions with other galaxies.

Voids provide unique laboratories for studying both cosmology and galaxy 
physics.  Because their evolution is partly shaped by the expansion of the 
universe \citep[e.g.,][]{Lavaux12, Pisani15, Sahlen19}, voids are also useful 
for studying dark energy and provide precision constraints on cosmological 
models via the Alcock-Paczy{\'n}ski test \citep[e.g.,][]{Lavaux12, Sutter12b, 
Sutter14b, Hamaus16, Mao17a, Nadathur19b, Hamaus20}, as well as in neutrino 
studies and models of galaxy formation \citep{Peebles01, Constantin08, 
Sahlen19}.  In addition, voids are used in measurements of baryon acoustic 
oscillations via void-galaxy correlation functions \citep{Nadathur19b, Zhao20, 
Zhao22}.  Due to the extreme low density of void interiors, void galaxies rarely 
interact with other objects, in contrast to galaxies found in denser regions.  
As a result, they provide relatively undisturbed environments for studying 
galactic evolution.

While voids do have some universal properties, such as an underdense center 
($\lesssim5\%$ of the mean galaxy density) and a steep density gradient at their 
boundary \citep{Colberg08}, the definition of what is a void is rather vague.  
In part, this is because void morphologies and densities exhibit significant 
variability.  Due to the ambiguity, a number of different void finding 
algorithms exist to identify underdense regions.

When making conclusions based on different void catalogs, care must be taken to 
untangle physical effects from the details of how voids are defined.  The 
Aspen-Amsterdam void finder comparison project \citep{Colberg08} performed a 
preliminary comparison between 13 void-finding algorithms, examining the 
distributions of the different algorithms' void properties.  A recent comparison 
of the void-finding algorithms in the Void Analysis Software Toolkit 
\citep[\VAST;][]{VAST} by \cite{Veyrat23} on their ability to recover regions of 
the large-scale structure whose dynamics match those expected for voids found 
that the different algorithms do identify different dynamical regions of the 
universe.  However, the impact of the choice of void-finding algorithms on 
either cosmology or galaxy science has not been systematically studied.  
\cite{Muldrew12} used mock galaxy catalogs to study how various definitions of 
the large-scale environment change the inferred effect of the environment on 
galaxy evolution.  In this work, we focus on the impact of the void-finding 
algorithms on inferred galaxy properties by comparing the properties of the 
galaxies inside voids defined by different algorithms, but it does not focus on 
void definitions or void algorithms.  This study allows us to understand how 
conclusions about the effects of the void environment depend on the choice of 
void-finding algorithm.

\section{Expected Galaxy Evolution in Voids}\label{sec:expectations}

N-body simulations and semi-analytic models of $\Lambda$CDM cosmology predict 
that void galaxies are retarded in their star formation relative to galaxies in 
denser regions \citep{Cen11}.  Numerous studies have been conducted to test this 
with observations; investigating the influence of the void environment on the 
galactic mass function \citep[e.g.,][]{Hoyle05, Croton05, Moorman15}, on galaxy 
photometry \citep[e.g.,][]{Grogin00, Rojas04, Rojas05, Patiri06, Park07, 
vonBendaBeckmann08, Hoyle12}, on the star formation rate 
\citep[e.g.,][]{Rojas05, Moorman16, Beygu16}, and on chemical composition 
\citep[e.g.,][]{Kreckel12, Douglass17b}.  The vast majority of these 
observational studies confirm that galaxies in voids are fainter and bluer than 
galaxies in denser regions.  This is likely a result of the retention of gas by 
void galaxies as they evolve in an environment with fewer interactions such as 
mergers, tidal stripping, ram-pressure stripping, etc., compared to galaxies in 
denser regions \citep{Cen11}.

We therefore expect void galaxies to be fainter, bluer, to have lower mass, and 
to have higher (specific) star formation rates regardless of the void-finding 
method used in the analysis.  We test this by comparing the properties of void 
galaxies as classified using two popular void-finding algorithms.  To achieve 
high statistical significance, we use a large sample of galaxies from the Sloan 
Digital Sky Survey Data Release 7 \citep[SDSS~DR7,][]{SDSS7}, which to our 
knowledge is the largest galaxy sample used to study properties of void galaxies 
to date. The comparisons shed light on the impact of different void-finding 
algorithms in galactic classification.

\section{Galaxy Catalog: SDSS DR7}\label{sec:data_set}

SDSS~DR7 \citep{SDSS7} is the final public data release from SDSS-II and was 
designed to collect spectra of a magnitude-complete galaxy sample, observing 
everything brighter than a Petrosian $r$-band magnitude $m_r < 17.77$ 
\citep{Strauss02}.  This final catalog includes 11,663 deg$^2$ of imaging data 
and five-band photometry for 357 million objects \citep{Lupton01}.  It also has 
complete spectroscopy with a resolution $\lambda/\Delta \lambda \sim 1800$ in 
the observed wavelength range 3800{\AA}--9200{\AA} \citep{Blanton03a} covering a 
contiguous 9380 deg$^2$ of the Northern Galactic Cap.  In total, the SDSS~DR7 
spectroscopic sample contains 1.6 million spectra including 930,000 galaxies, 
120,000 quasars, and 460,000 stars.

We use SDSS~DR7 as reprocessed and presented in the NASA-Sloan 
Atlas\footnote{\url{http://www.nsatlas.org/}} (NSA, version 1.0.1), a catalog of 
images and parameters derived from the SDSS imaging \citep{Blanton11}.  The NSA 
provides $K$-corrected absolute magnitudes, colors, and stellar masses.  Star 
formation rates and specific star formation rates are taken from the MPA-JHU 
value-added galaxy 
catalog\footnote{\url{https://wwwmpa.mpa-garching.mpg.de/SDSS/DR7/}} based on 
the method described in \cite{Brinchmann04}.  We assume a flat $\Lambda$CDM 
cosmology with $\Omega_M = 0.315$ and $H_0 = 100h$~km~s$^{-1}$~Mpc$^{-1}$.

\section{The Void Analysis Software Toolkit}\label{sec:vast}

\begin{figure*}
  \includegraphics[height=2.05in]{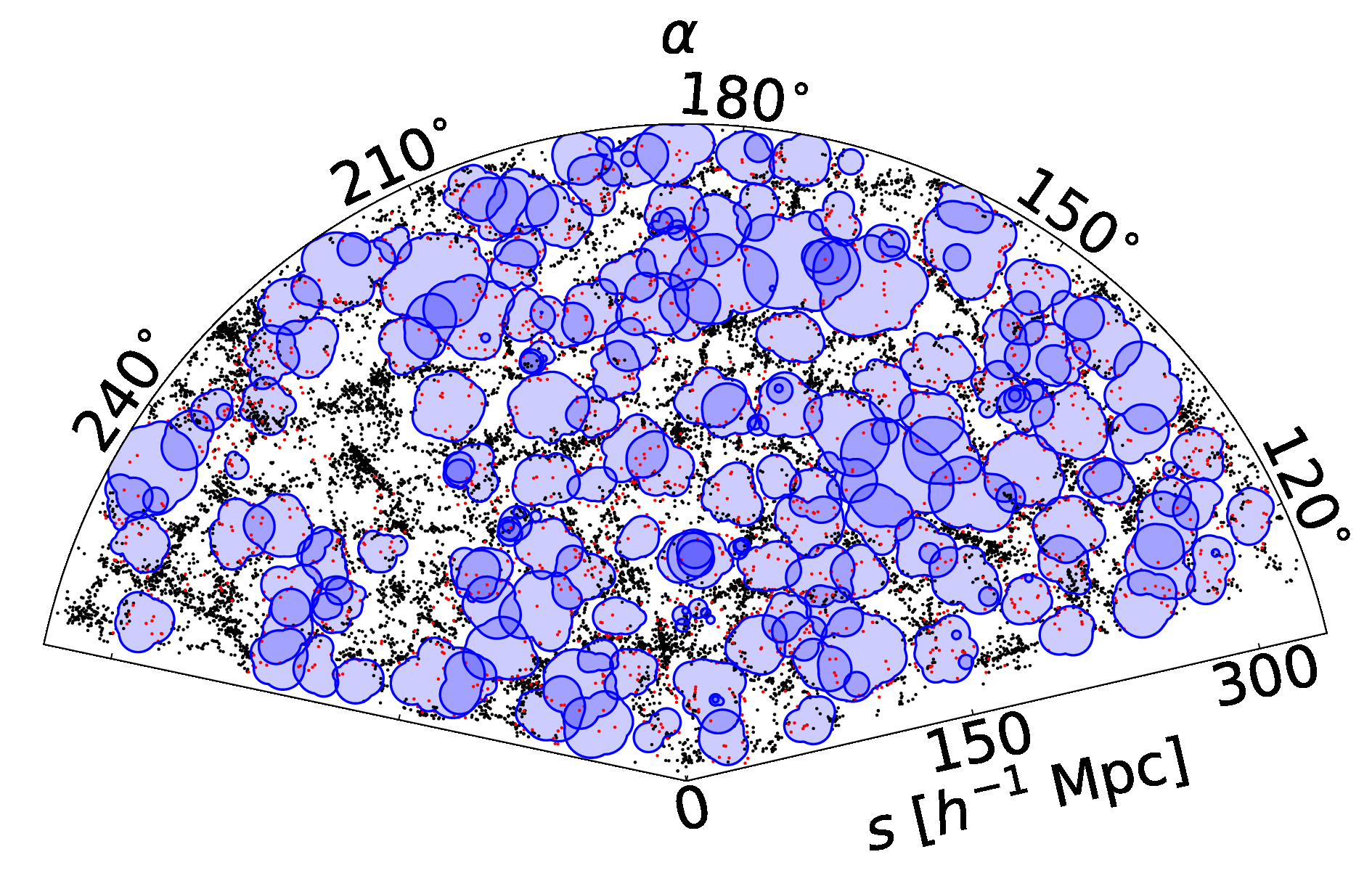}
  \includegraphics[height=2.05in]{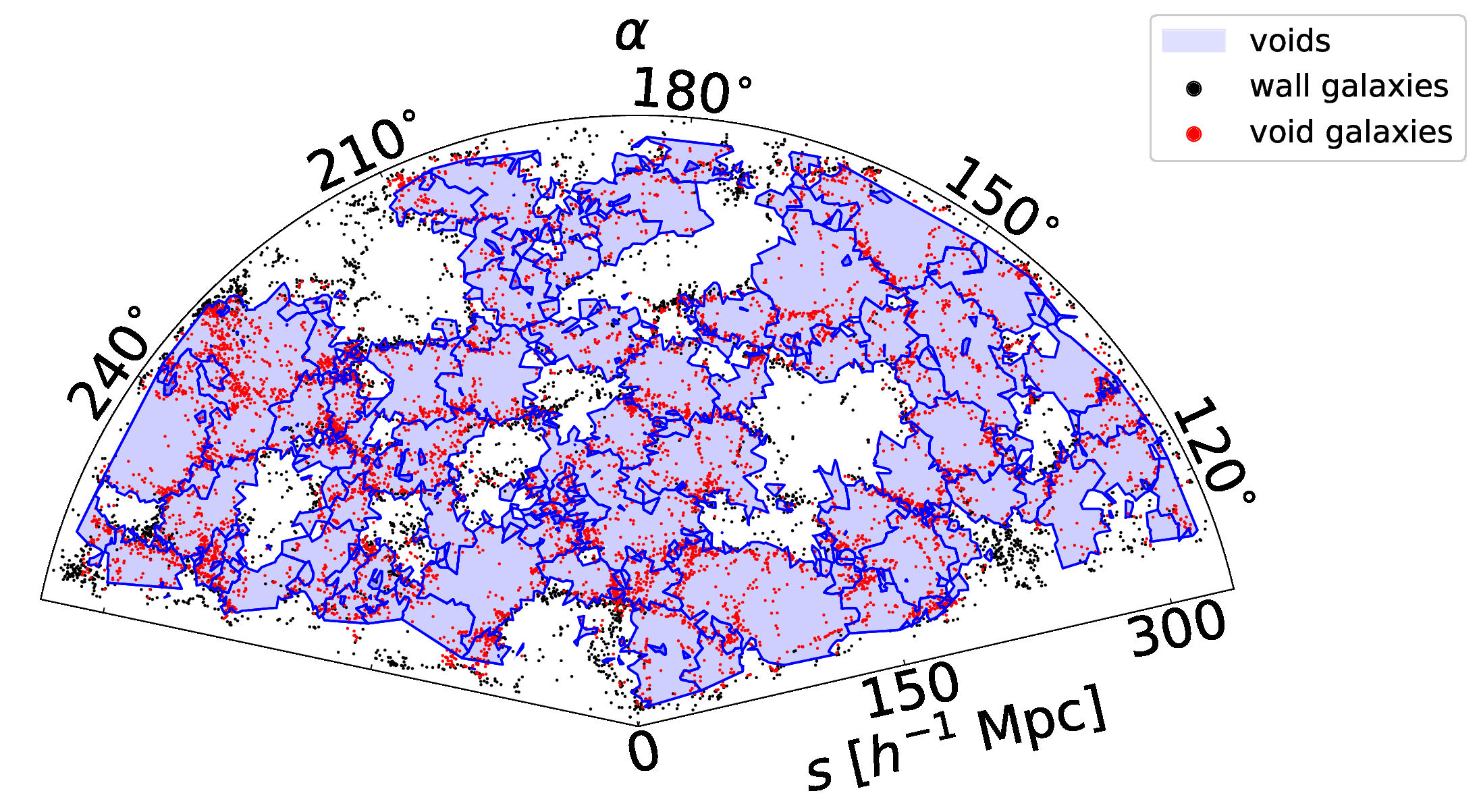}
  \caption{Results from \VAST's \VF (left) and \VV (right) for a 10\hMpc-thick 
  slice centered on declination $\delta=32^\circ$ of the main SDSS~DR7 galaxy 
  catalog.  The intersection of the midplane of the declination range and the 
  voids are shown in blue; galaxies that fall within a void are colored in red, 
  while those that do not are shown in black.  Projection effects can produce 
  void galaxies appearing to fall outside a void, and some wall galaxies 
  appearing to fall within a void.}
  \label{fig:DR7_slice}
\end{figure*}

We use the void catalogs presented in \cite{Douglass23}.  The catalogs were 
produced using a volume-limited sample of SDSS~DR7 ($M_r \leq -20$, $z \leq 
0.114$) with two popular void-finding algorithms implemented in the Void 
Analysis Software Toolkit \citep[\VAST;][]{VAST}: \VF and \VV.  \VF 
\citep{ElAd97} is based on the growth and merging of nearly empty spheres, 
motivated by the expectation that voids are spherical to first order.  \VV 
(\underline{V}oronoi \underline{V}oids) is a Python 3 implementation of the 
\texttt{ZOBOV} \citep[\underline{ZO}nes \underline{B}ordering \underline{O}n 
\underline{V}oidness;][]{Neyrinck08} algorithm, which defines voids based on 
local minima in the galactic density field.  An example of the results from the 
\VAST void-finding algorithms applied to SDSS~DR7 are shown in 
Figure~\ref{fig:DR7_slice}.  Any galaxy which falls within one of the void 
regions defined by a given algorithm is considered a void galaxy; all others are 
classified as wall galaxies.

\subsection{\VF}\label{sec:vf}

\VF begins by identifying and removing all field galaxies, defined as those 
galaxies with their third nearest neighbor farther than 5.33\hMpc.  The 
remaining (wall) galaxies are placed on a three dimensional grid with side 
lengths of 5\hMpc.  Spheres are grown from the center of each empty grid cell 
and expand until they are bounded by four galaxies on their surface.  Each void 
is defined as a union of these spheres, where the largest sphere within a void 
is that void's maximal sphere.  Maximal spheres are not permitted to overlap 
each other by more than 10\% of their volume, and they have a minimum radius of 
10\hMpc.  All other spheres of a given void must overlap only their void's 
maximal sphere by at least 50\% of their volumes.  For more details, see 
Section~\ref{sec:VF_appendix} and \cite{Douglass23}.

\subsection{Voronoi Voids (\VV)}\label{sec:zobov}

\VV first produces a 3D Voronoi tessellation of the input galaxy catalog.  The 
Voronoi cells are combined into ``zones'' using a watershed segmentation.  Zones 
are then merged into voids if the least-dense pair of adjacent cells between two 
zones is below a given maximum ``linking density.''  Here, we define the 
zone-linking density threshold as $0.2\bar{\rho}$, where $\bar{\rho}$ is the 
mean density of the SDSS~DR7 volume-limited galaxy catalog, similar to the 
threshold used in \texttt{VIDE} \citep{Sutter15}.  In addition, we prune our \VV 
catalog to remove all voids with an effective radius smaller than 10\hMpc.  For 
more details, see Section~\ref{sec:V2_appendix} and \cite{Douglass23}.

Unlike \VF, which assumes that the voids are spherical to first order, \VV makes 
no assumptions regarding void shape.  In addition, \VV depends on fewer tunable 
parameters than \VF, making it more survey-agnostic.

\section{Comparison of void galaxy properties between algorithms}\label{sec:influence_results}

\input{t1}

We investigate the influence of the void environment on inferred galaxy 
properties, where the environmental classification (void or wall) is determined 
from each of the void catalogs described in Section~\ref{sec:vast}.  To study 
how the inferred properties of void galaxies depend on the choice of 
void-finding algorithm, we plot the distribution of each property for the void 
and wall galaxy populations as defined by \VF and \VV.  We report the mean, 
median, and corresponding difference between the means and medians of the void 
and wall galaxy distributions for each property and algorithm in 
Table~\ref{tab:base_model}.

Two Bayesian analysis models are used to quantify the statistical significance 
of differences between the inferred properties of void and wall galaxies.  
Properties with unimodal distributions (luminosity and stellar mass) are modeled 
as a mixture of two skew normal distributions; properties with bimodal 
distributions (color and (s)SFR) are modeled as a mixture of three skew normal 
distributions.  We then compute the Poisson likelihoods that the void and wall 
galaxy properties are drawn from the same parent distribution and from different 
parent populations.  The likelihood ratio of the one- and two-parent models is 
used to construct a Bayes factor, $B_{12}$, to discriminate between the models 
(see Appendix~\ref{sec:BF_appendix} for details). 

For all properties and void-finding algorithms, the Bayes factor gives decisive 
evidence in favor of the two-parent model.  However, the differences between 
void and wall galaxies are substantially less significant for \VV than \VF.  
The Bayes factors for all galaxy properties and void-finding algorithms are 
listed in the rightmost column of Table~\ref{tab:base_model}.  In the following 
subsections, we describe differences between the stellar mass, color, 
luminosity, and (specific) star formation rate of void and wall galaxies using 
\VV and \VF for void galaxy classification.


\subsection{Stellar Mass}\label{ssec:results_stellar_mass}

\begin{figure*}
  \includegraphics[width=0.49\textwidth]{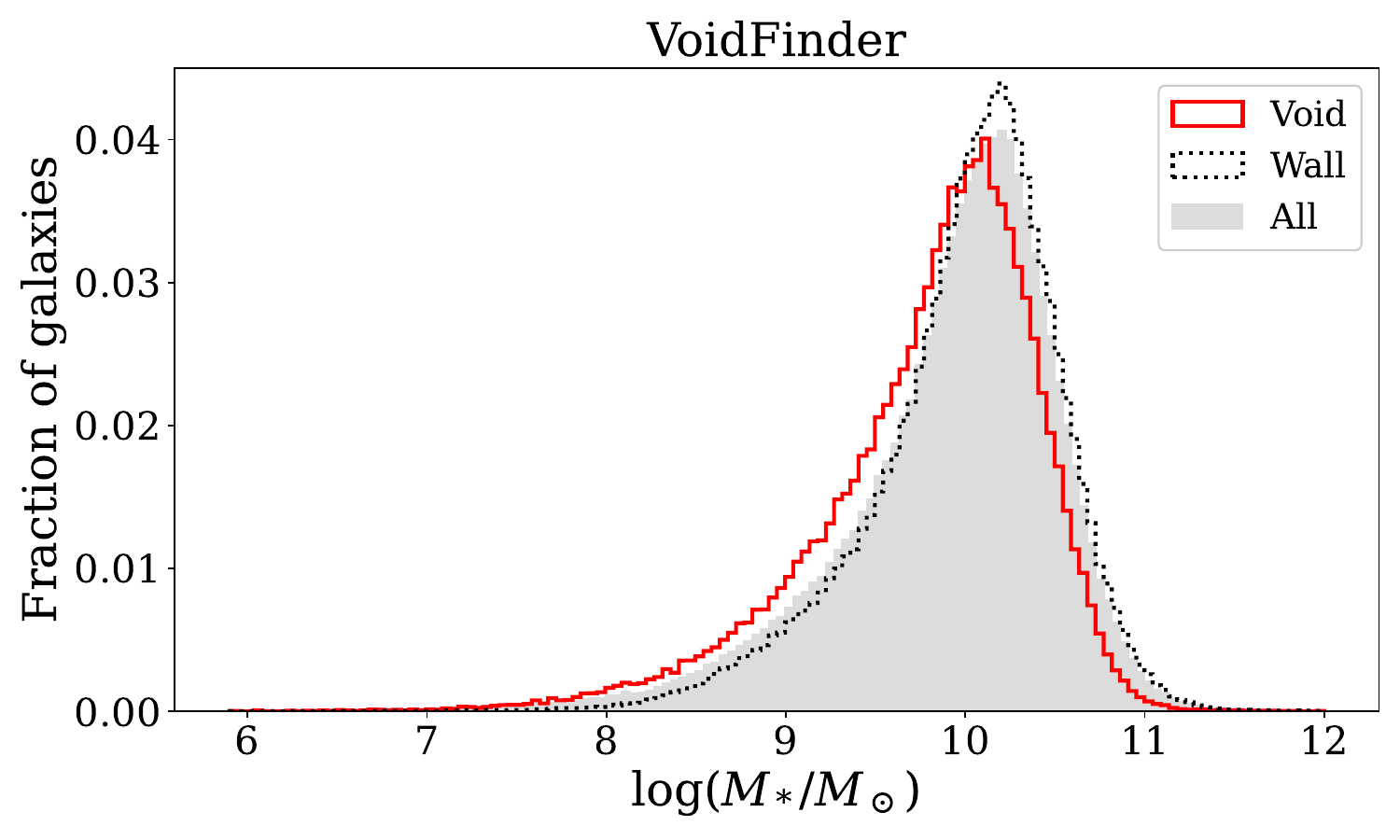}
  \includegraphics[width=0.49\textwidth]{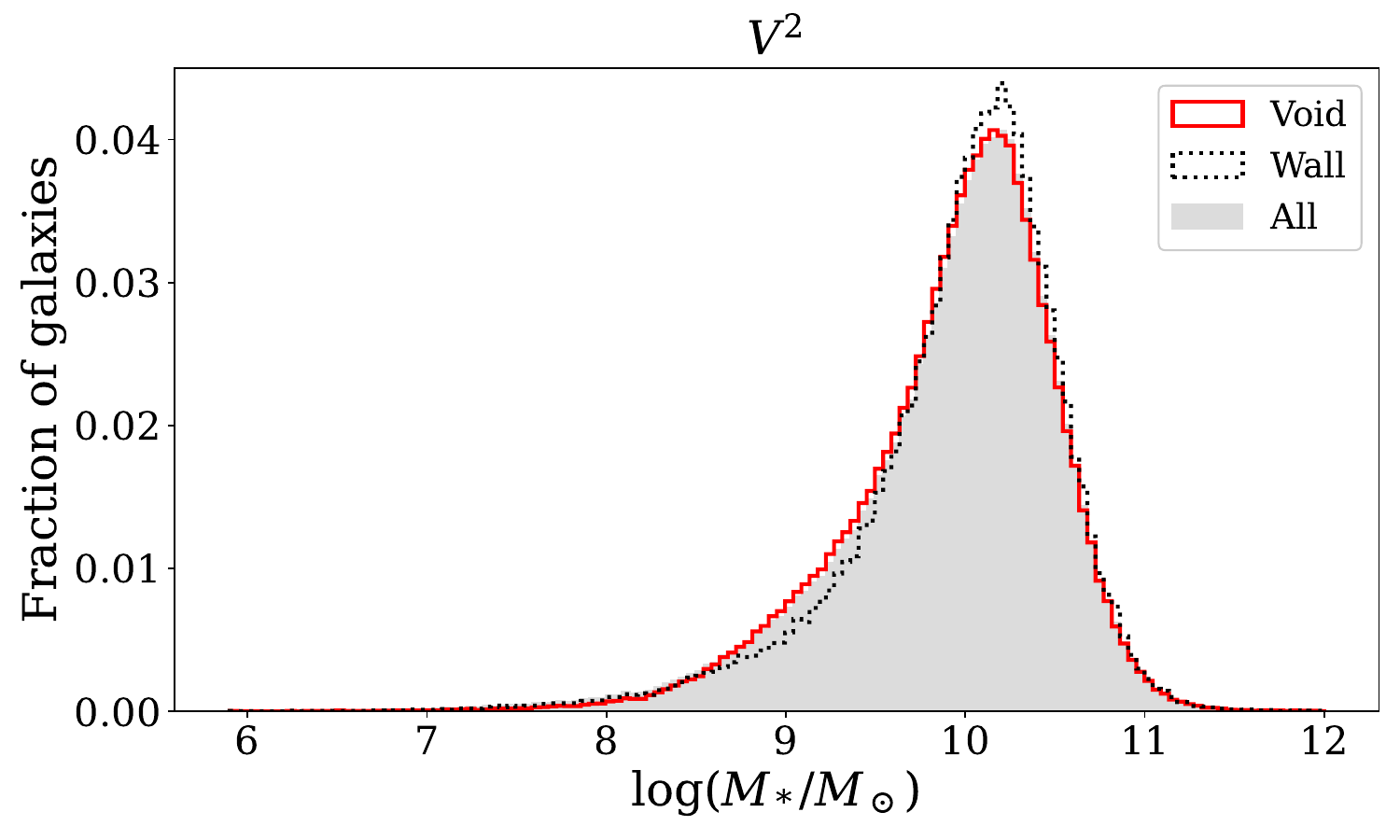}
  \caption{Stellar mass distribution of galaxies, separated into void and wall 
  environments according to the \VF (left) and \VV (right) void catalogs.  The 
  distribution of all galaxies is shaded in gray.  There is a shift towards 
  lower stellar mass for void galaxies relative to wall galaxies with both void 
  classifications, although \VV's is very slight.}
  \label{fig:Mstar}
\end{figure*}

We study the distribution of stellar masses of void galaxies compared to the 
distribution of galaxies in denser regions, as shown in Figure~\ref{fig:Mstar}.  
Void galaxies are expected to have systematically lower stellar masses than 
galaxies in denser regions because galaxy interactions, much more rare in voids, 
are expected to enhance star formation.  Both the distributions shown in 
Figure~\ref{fig:Mstar} and the statistics reported in Table~\ref{tab:base_model} 
show that the stellar mass distribution for void galaxies is statistically 
different than the wall galaxy sample when either void-finding algorithm is used 
for void classification.  When the void environment is defined with \VF, we see 
this difference manifest as a relatively large shift towards lower stellar 
masses in the voids.  With \VV, though, we find that the difference between the 
void and wall populations is very slight and appears to be a result of a slight 
narrowing of the stellar mass distribution in walls, as the void galaxy 
distribution appears to match the distribution of all galaxies.

\subsection{Luminosity}\label{ssec:abs_magnitude_results}

\begin{figure*}
  \includegraphics[width=0.49\textwidth]{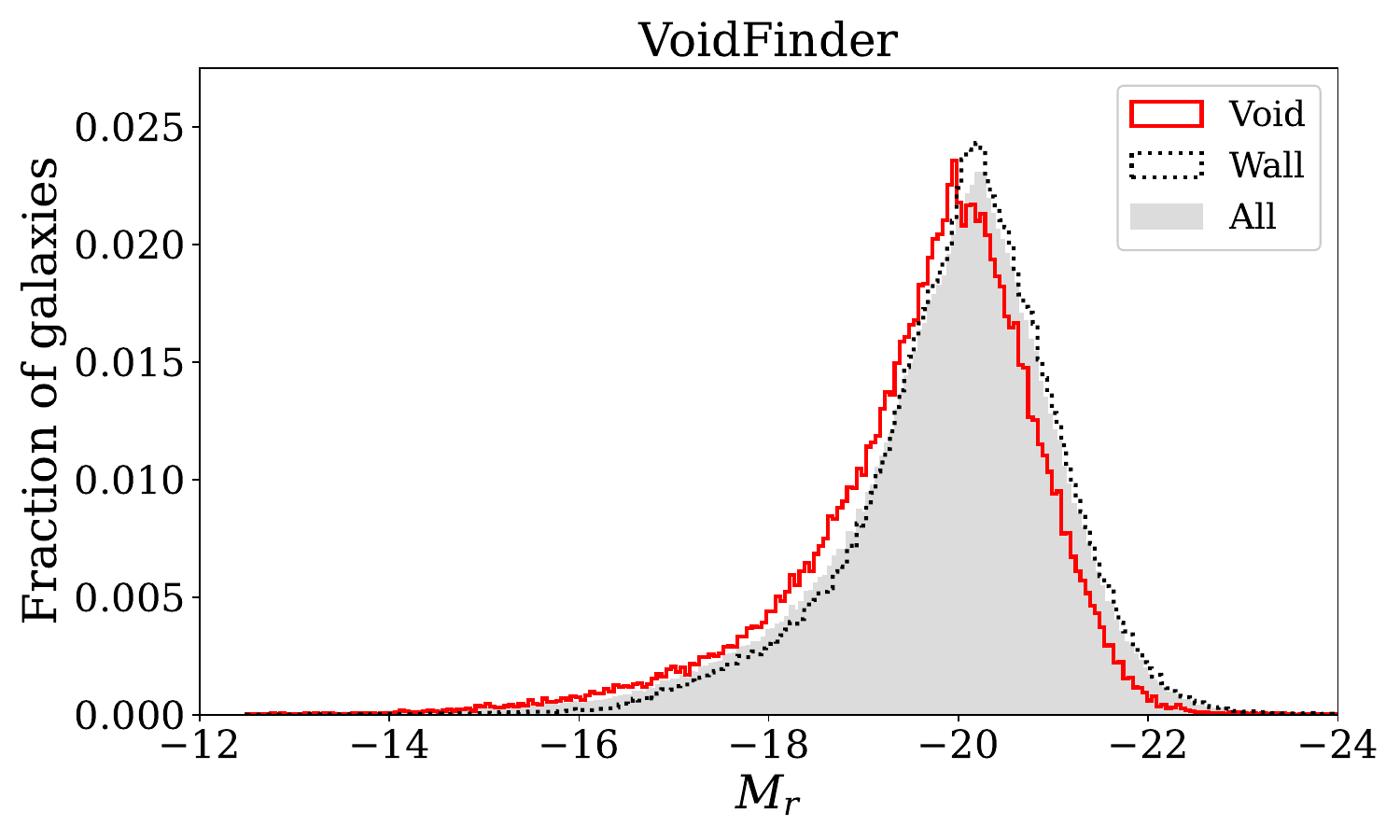}
  \includegraphics[width=0.49\textwidth]{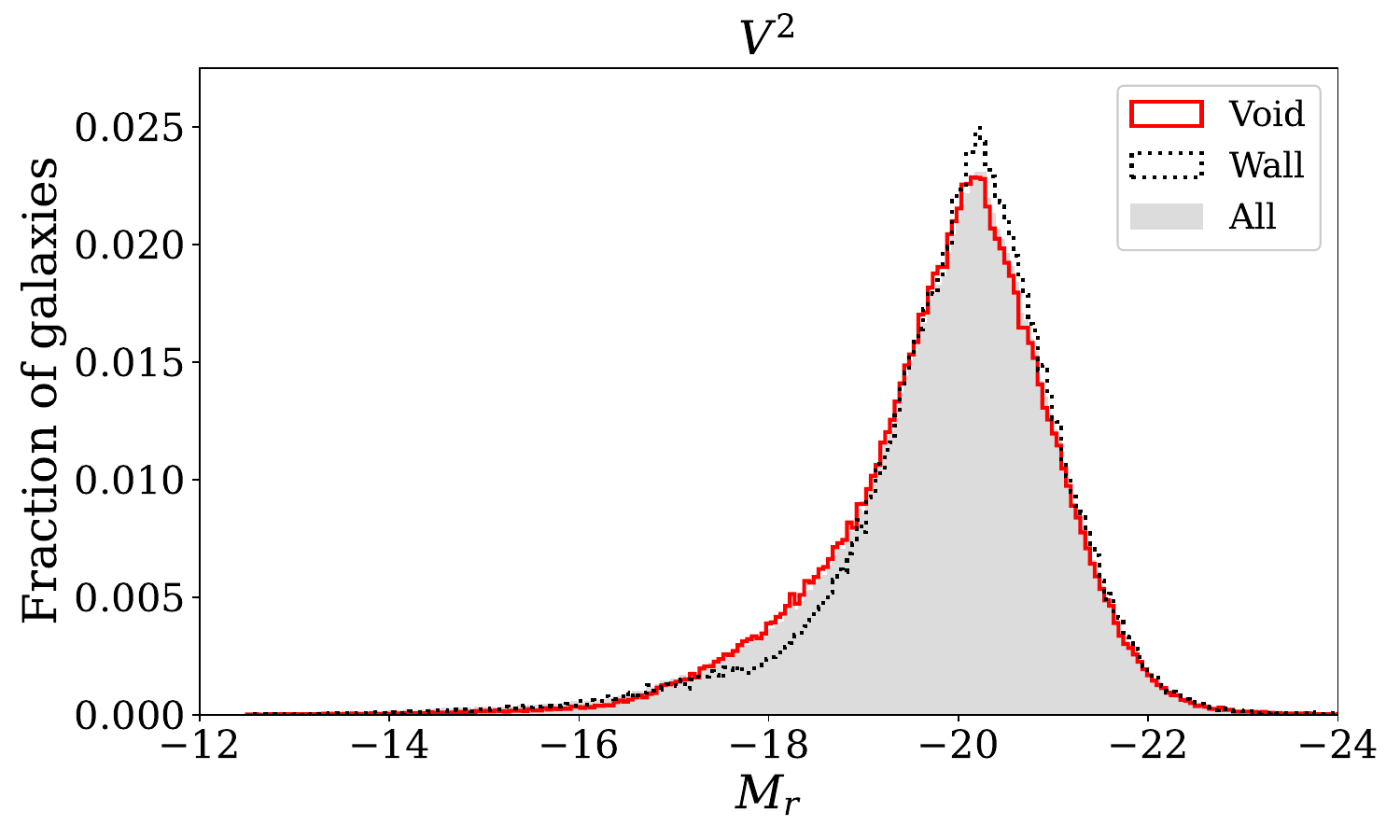}
  \caption{Luminosity distribution of galaxies separated according to void and 
  wall environments by \VF (left) and \VV (right).  The distribution for all 
  galaxies is shaded in gray.  There is a shift towards fainter luminosities in 
  the distribution of void galaxies relative to wall galaxies in \VF's 
  classification, while it appears that the wall galaxies are shifted towards 
  slightly brighter luminosities with \VV's classification.}
  \label{fig:abs_mag}
\end{figure*}

A galaxy's luminosity is closely linked to its stellar mass, so we expect void 
galaxies to be systematically fainter than galaxies in denser regions.  As seen 
on the left in Figure~\ref{fig:abs_mag}, the classification by \VF results in a 
shift towards fainter luminosities for void galaxies than galaxies in denser 
environments.  This matches the trend observed in the stellar mass, shown in 
Figure~\ref{fig:Mstar}.  Using the \VV void catalog, though, we see very little 
difference between the void and wall galaxy luminosity distributions, as seen on 
the right in Figure~\ref{fig:abs_mag}; the void galaxies as defined by \VV have 
nearly the same luminosities as its wall galaxies, with possibly a slight 
thinning of the wall population.  The Bayes factors (listed in 
Table~\ref{tab:base_model}) corroborate these differences between the void and 
wall populations for both classifications.

\subsection{(specific) Star formation rate}\label{ssec:star_formation_results}

\begin{figure*}
  \includegraphics[width=0.49\textwidth]{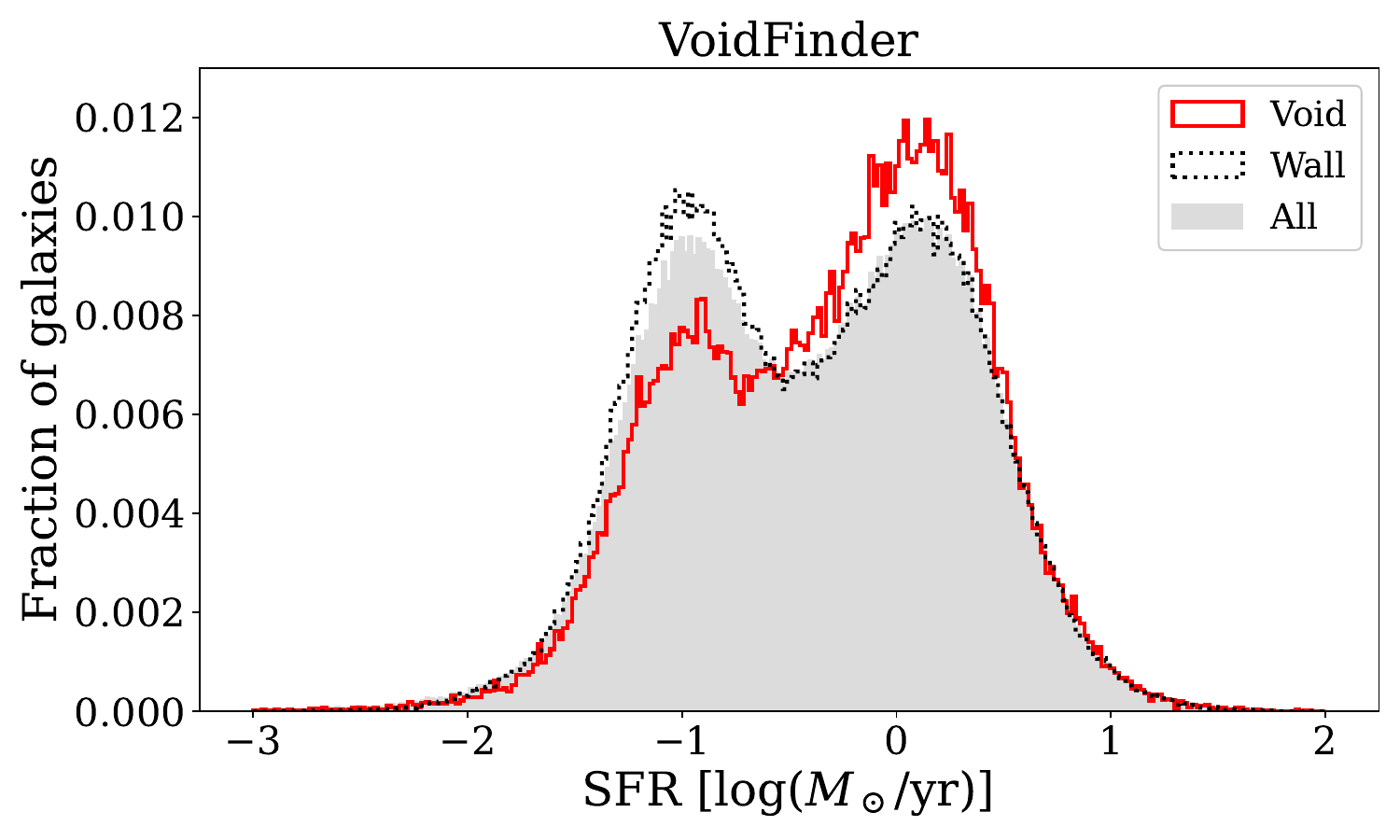}
  \includegraphics[width=0.49\textwidth]{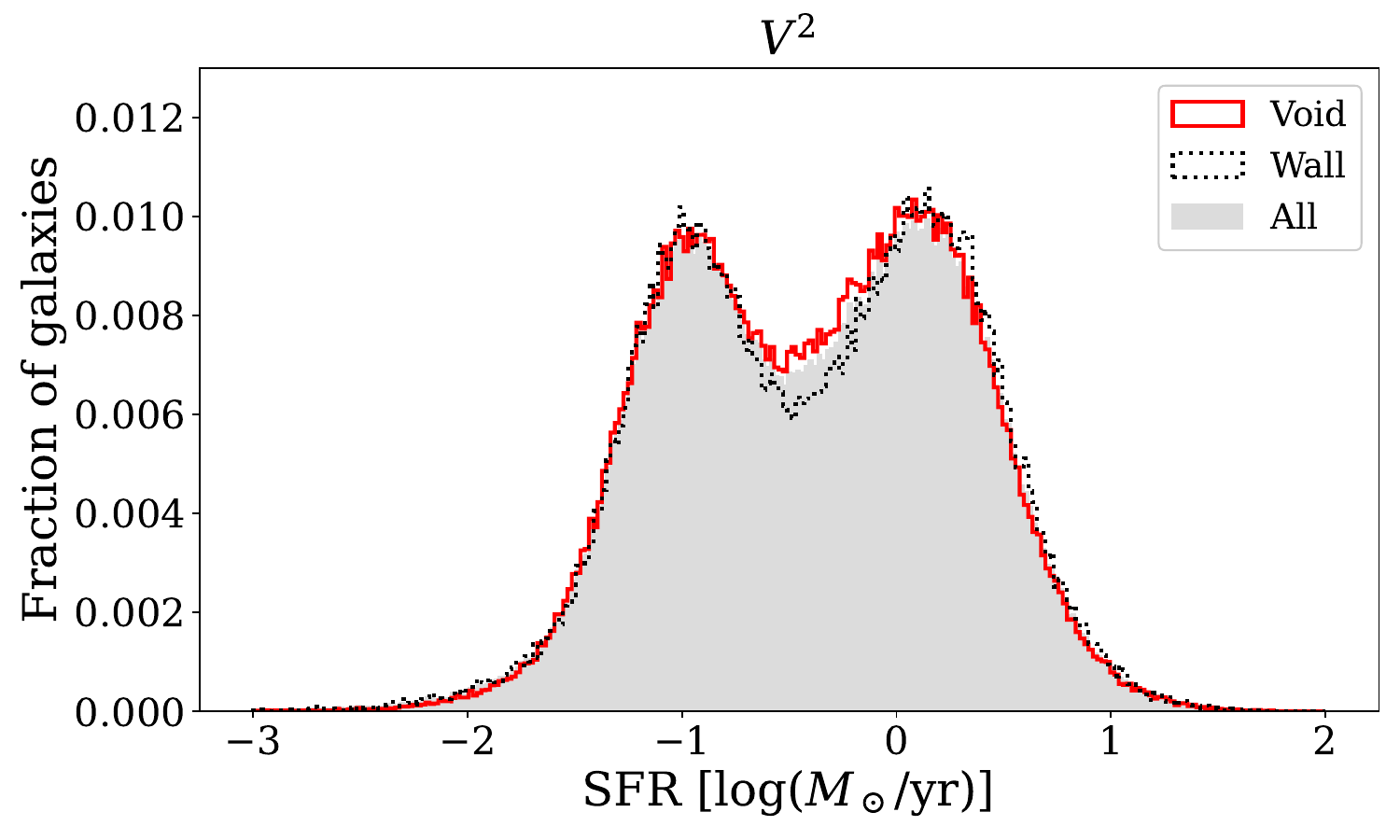}
  \includegraphics[width=0.49\textwidth]{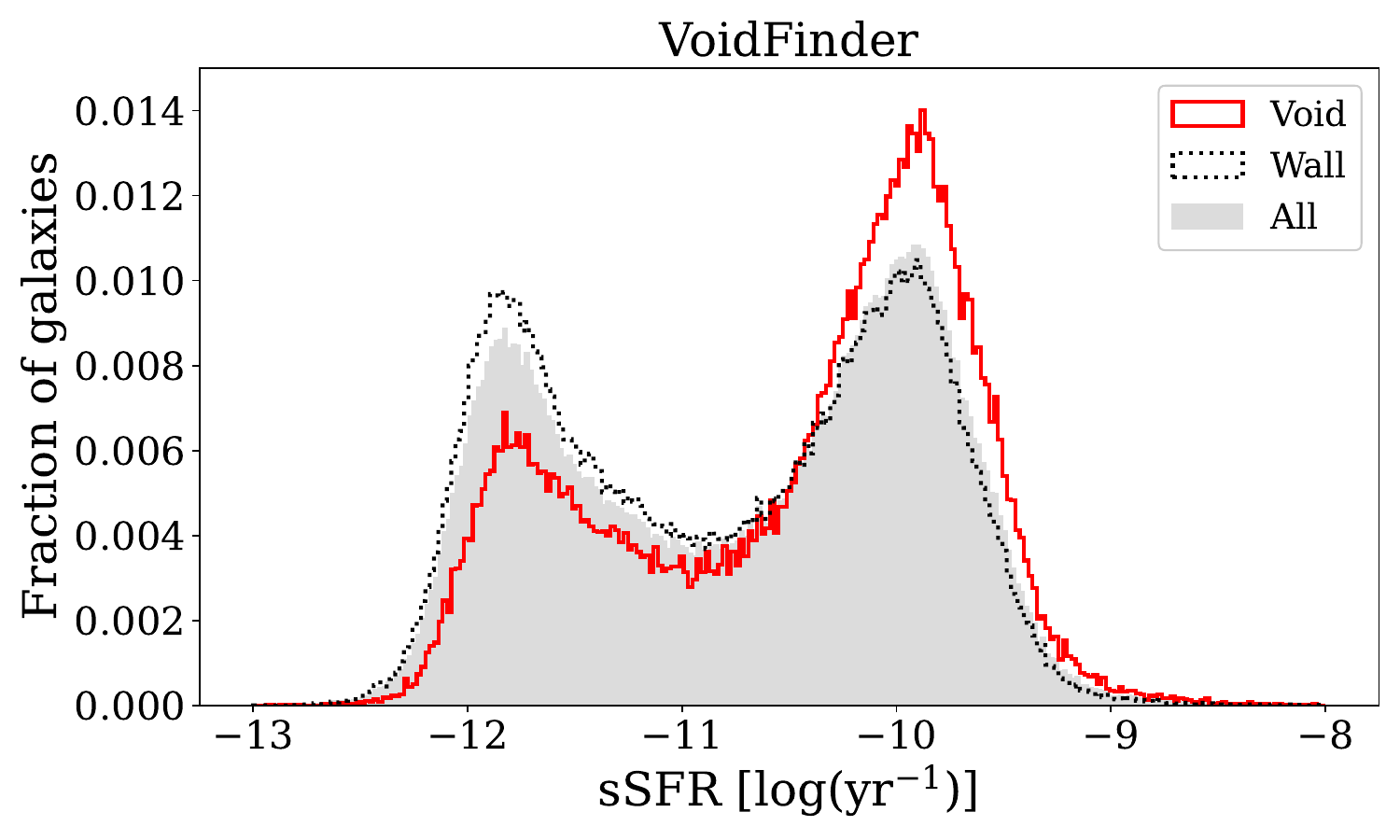}
  \includegraphics[width=0.49\textwidth]{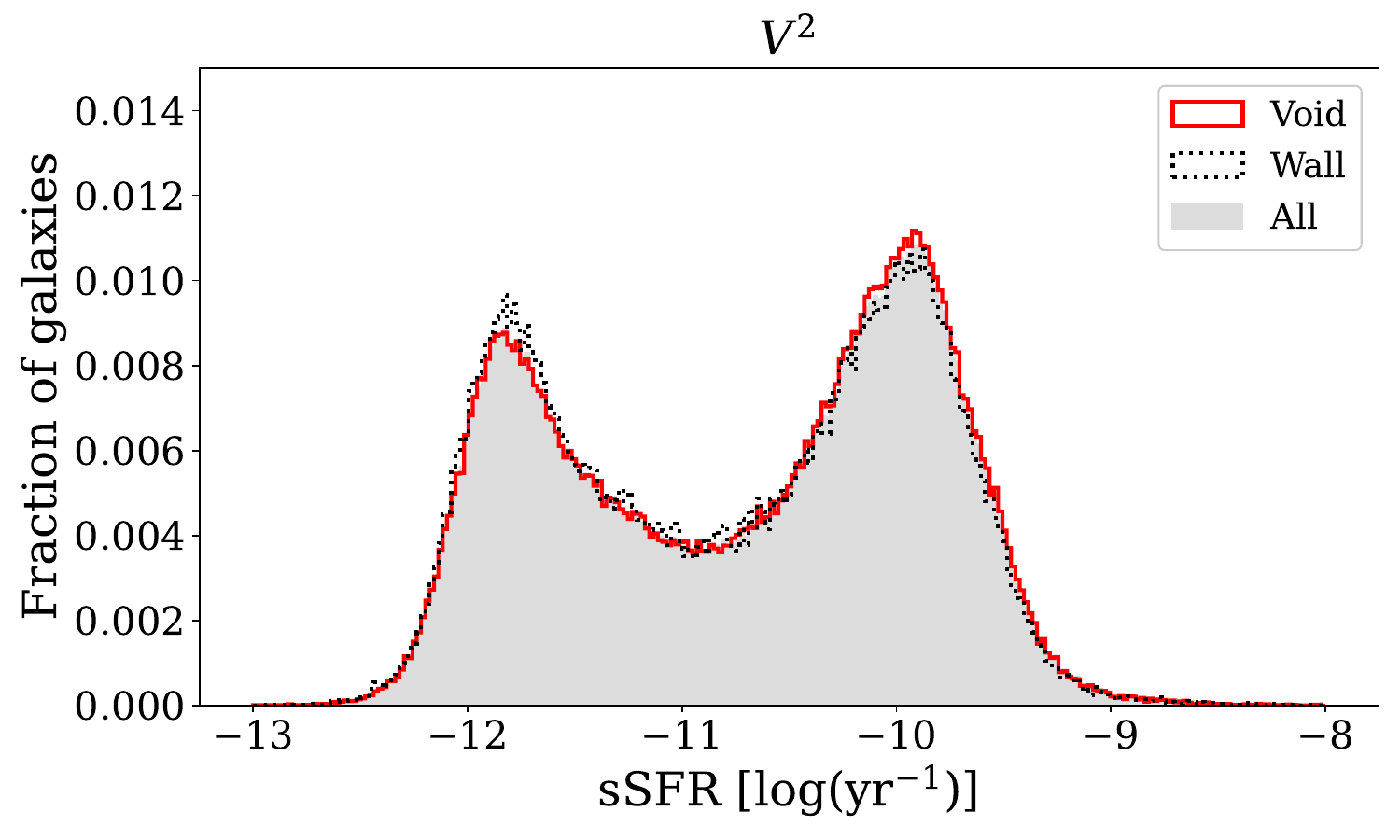}
  \caption{Star formation rate (SFR, top) and specific star formation rate 
  (sSFR, bottom) distributions of galaxies for void and wall environments as 
  classified by \VF (left) and \VV (right).  The distribution of all galaxies is 
  shown in gray for comparison.  When classified using \VF's void catalog, void 
  galaxies have systematically higher (specific) star formation rates relative 
  to wall galaxies; we observe no discernible difference between void and wall 
  galaxies when the \VV void catalog is used.}
  \label{fig:(s)SFR}
\end{figure*}

Void galaxies are expected to be undergoing more star formation today than 
galaxies in denser regions, due both to potential retarded star formation and 
their continued presence of cold gas reservoirs \citep{Cen11}.  Therefore, void 
galaxies are expected to have higher star formation rates (SFR) and specific 
star formation rates (sSFR; the star formation rate per unit mass) than galaxies 
in denser regions.  As shown in the left column of Figure~\ref{fig:(s)SFR}, we 
find that void galaxies have higher (s)SFR values relative to galaxies in denser 
environments when using the \VF void catalog for environmental classification.  
As seen on the right in Figure~\ref{fig:(s)SFR}, the distributions for \VV show 
no such trend in (s)SFR --- according to the \VV void catalog, both void and 
wall galaxies have the same (s)SFR.  

The Bayes factors (listed in Table~\ref{tab:base_model}) support the significant 
difference between the void and wall populations of \VF for the (s)SFR 
distributions.  However, the Bayes factors for the (s)SFR distributions for void 
and wall galaxies as classified by \VV also indicate that the two populations 
come from different parent distributions.  From the distributions shown on the 
right in Figure~\ref{fig:(s)SFR}, this is possibly due to a very slight shift in 
the \VV wall galaxy population towards lower (s)SFRs, not a shift in the void 
galaxy population towards higher (s)SFRs as seen with \VF.

\subsection{Color}\label{ssec:color_results}

\begin{figure*}
  \includegraphics[width=0.49\textwidth]{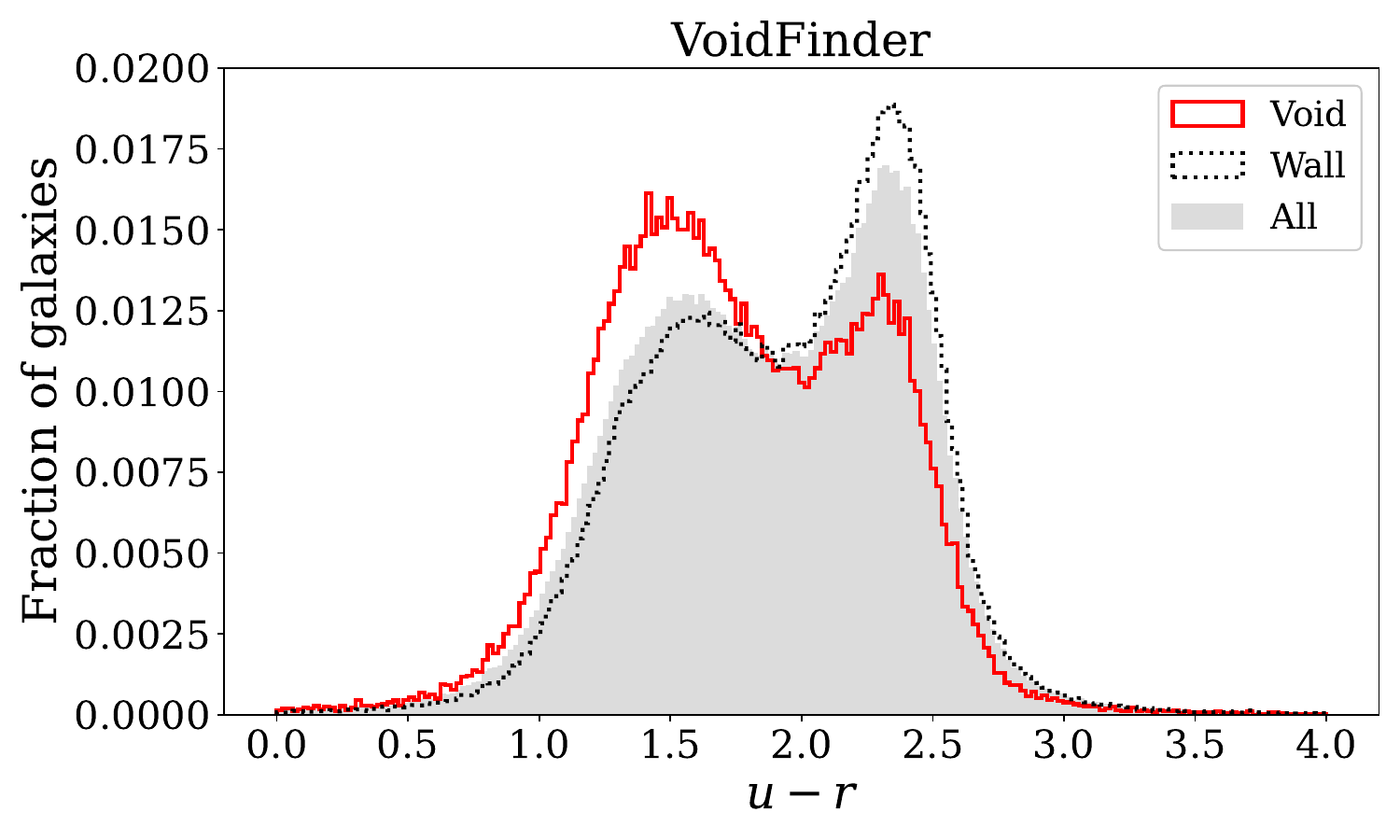}
  \includegraphics[width=0.49\textwidth]{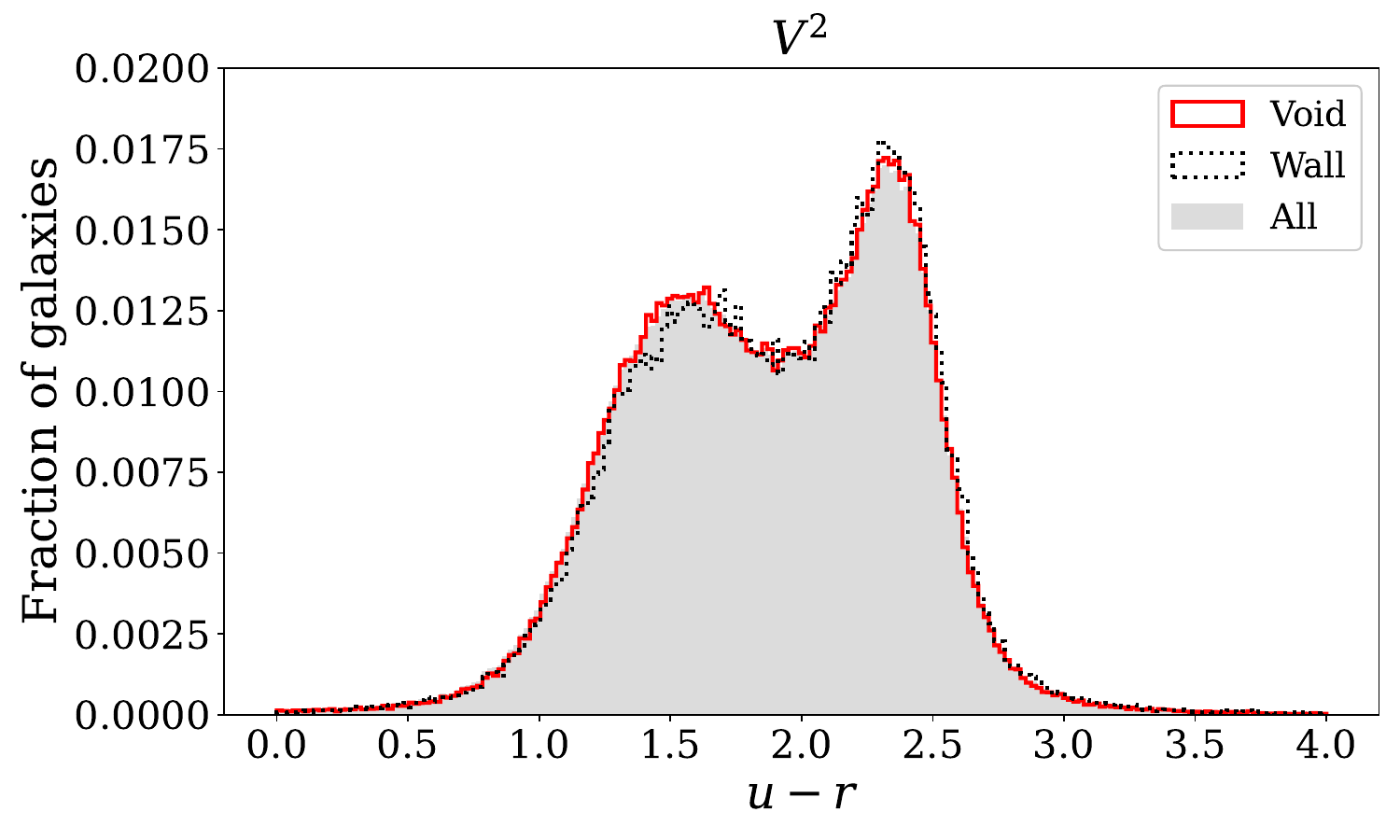}
  \includegraphics[width=0.49\textwidth]{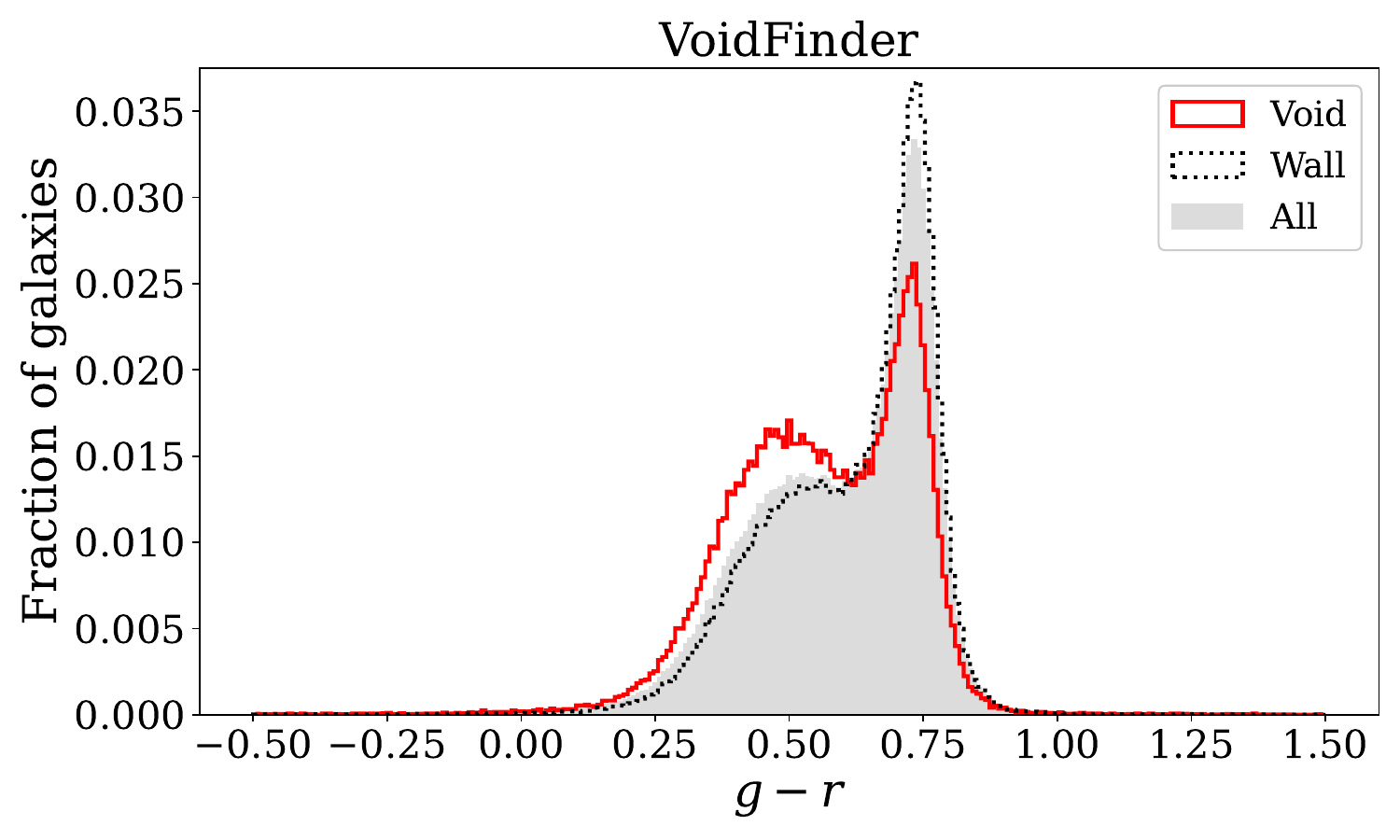}
  \includegraphics[width=0.49\textwidth]{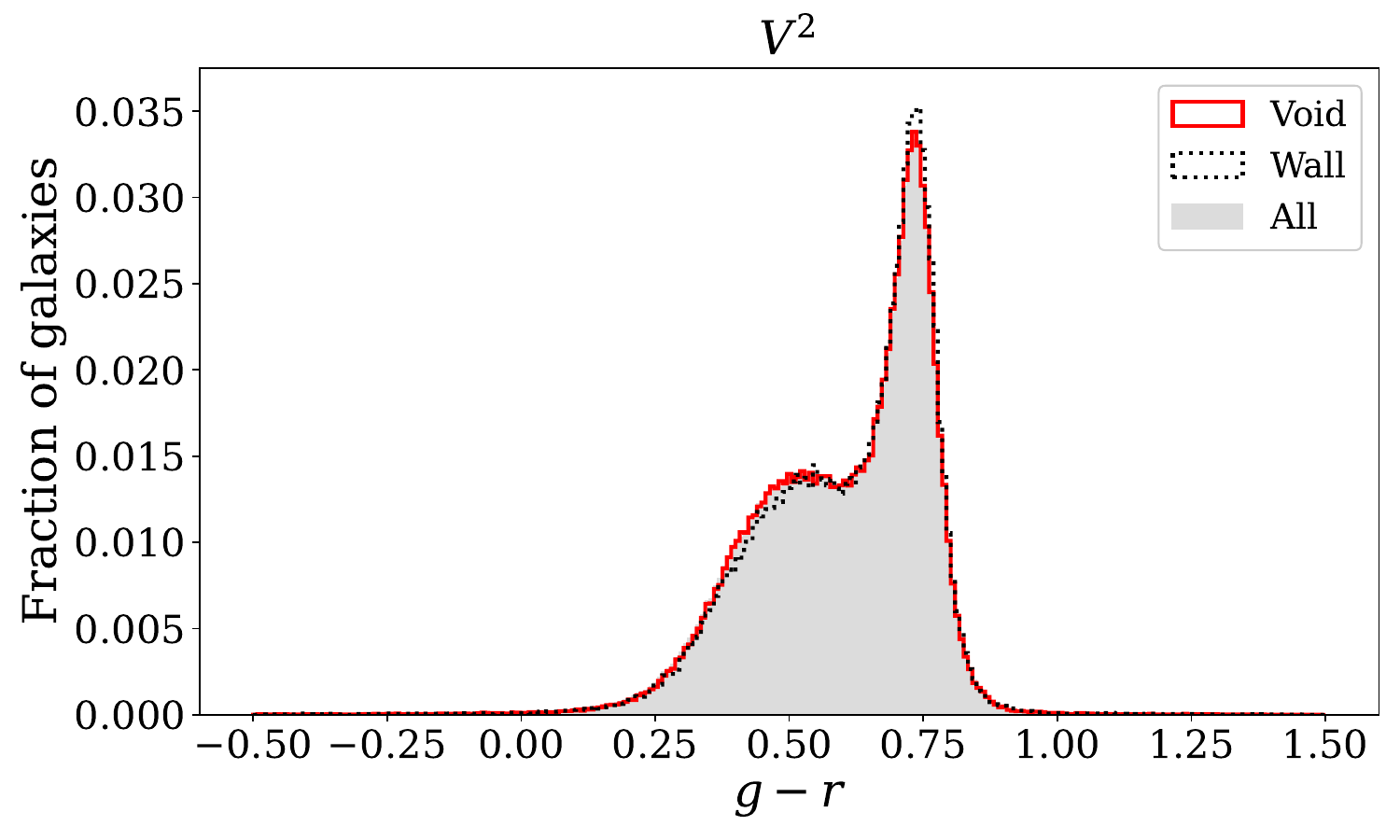}
  \caption{Color distribution, $u-r$ (top row) and $g-r$ (bottom row), separated 
  by their environment (void, wall) using \VF (left column) and \VV (right 
  column).  The distribution for all galaxies is given in gray.  There is an 
  apparent shift towards bluer color for void galaxies relative to wall galaxies 
  with \VF's classification but not \VV's.}
  \label{fig:color}
\end{figure*}

For galaxies in SDSS~DR7, we explore the distribution of the color of void 
galaxies relative to galaxies in denser regions as classified by both \VF and 
\VV, as shown in Figure~\ref{fig:color}.  A galaxy's color is determined by its 
stellar populations: bluer colors are a result of hot O and B stars, indicating 
that star formation has recently occurred, and red colors are indicative of an 
old stellar population.  With star formation expected to shift to later times in 
void regions, void galaxies are expected to be systematically bluer than 
galaxies in denser regions.  This is exactly the shift that we see in the \VF 
void galaxy population, shown on the left in Figure~\ref{fig:color} and 
reflected in the Bayes factor reported in Table~\ref{tab:base_model} comparing 
the \VF galaxy distributions.

The Bayes factors for the \VV color distributions are two orders of magnitude 
smaller than those of \VF.  Combined with the distributions shown on the right 
in Figure~\ref{fig:color}, we conclude that the differences quantified by the 
Bayes factors in the \VV distributions are actually a result of a slight 
increase in redder galaxies in the wall population, as evidenced by the slight 
increase in amplitude of the right-hand peak and slight decrease in the 
left-hand peak of the \VV wall distributions compared to the void and overall 
distributions.

\section{Discussion}\label{sec:discussion}

In Section~\ref{sec:influence_results}, we compare the distributions of a range 
of galaxy properties for galaxies in voids and galaxies in denser regions using 
two different definitions of the void environment.  With \VF, we find that void 
galaxies are systematically fainter, bluer, less massive, and have considerably 
higher star formation rates than galaxies in denser regions.  With \VV, though, 
we see very little difference between the properties of void galaxies and the 
properties of galaxies in denser regions.

\subsection{Impact of the void-finding algorithm on galaxy classification}

The $\Lambda$CDM cosmology predicts void galaxies to be retarded in their star 
formation relative to galaxies in denser regions since they experience fewer 
interactions (which often initiate star formation episodes) compared to galaxies 
in denser regions.  Additionally, they are able to retain their gas as they 
evolve in these extremely underdense environments \citep{Cen11}, allowing them 
to continue forming stars up to the present epoch.  The void galaxy properties 
with \VF used to define void regions are therefore consistent with the simulated 
physical properties of void galaxies \citep{Benson03a, Cen11, Habouzit20, 
Peper21} and with previous observations: void galaxies are fainter 
\citep[e.g,][]{Croton05, Hoyle05, Sorrentino06, Kreckel12, Moorman15}, less 
massive \citep[e.g.,][]{Jung14, Alpaslan16, Beygu17, Martizzi20}, bluer 
\citep[e.g.,][]{Grogin99, Rojas04, Croton05, Patiri06, Sorrentino06, 
vonBendaBeckmann08, Hoyle12, Kreckel12}, and have higher (s)SFR 
\citep[e.g.,][]{Rojas05, vonBendaBeckmann08, Cybulski14, Liu15, Beygu16, 
Moorman16}.

However, we do not see these expected traits for galaxies in voids defined by 
\VV.  To better understand the source of this discrepancy, we investigate the 
\VV void galaxies in more detail.  We find that 23.4\% of \VF wall galaxies are 
classified as void galaxies by \VV, and 40.5\% of \VV void galaxies are 
classified as wall galaxies by \VF.  As described in Section~\ref{sec:vast}, \VF 
voids are found by expanding spheres until they reach a relatively bright galaxy 
that is not isolated, while \VV defines voids by connecting Voronoi cells until 
they reach local density maxima.  Since these density maxima are likely to be 
close to the centers of the walls/filaments rather than their edges (where the 
brighter non-isolated galaxies that \VF stops at are located), \VV likely 
extends further into what \VF considers a wall.  This leads to \VV classifying 
many galaxies as void that are, instead, classified as wall galaxies by \VF.
 
Considering how the algorithms define the edges of voids and how \VF's results 
are consistent with expectations, we conclude that the void environment of \VV 
is contaminated by wall galaxies, similar to that shown by \cite{Florez21} and 
\cite{Veyrat23}.  \VV's void environment is therefore expected to behave more 
like the general distribution of galaxies.  Our observations agree with the 
simulation results of \cite{Veyrat23}: \VF provides a better delineation between 
the dynamically-distinct void regions and the surrounding wall structures, 
leading to more reasonable physical conclusions regarding the properties of void 
galaxies.  Unless additional restrictions are implemented, this contamination 
results in the void-wall classification from \VV, and thus ZOBOV-like 
algorithms, to not be suitable for void environment studies of galaxy formation 
and evolution.

\subsection{Galaxy properties as a function of depth within \VV voids}

\begin{figure}
  \centering
  \includegraphics[width=0.5\textwidth]{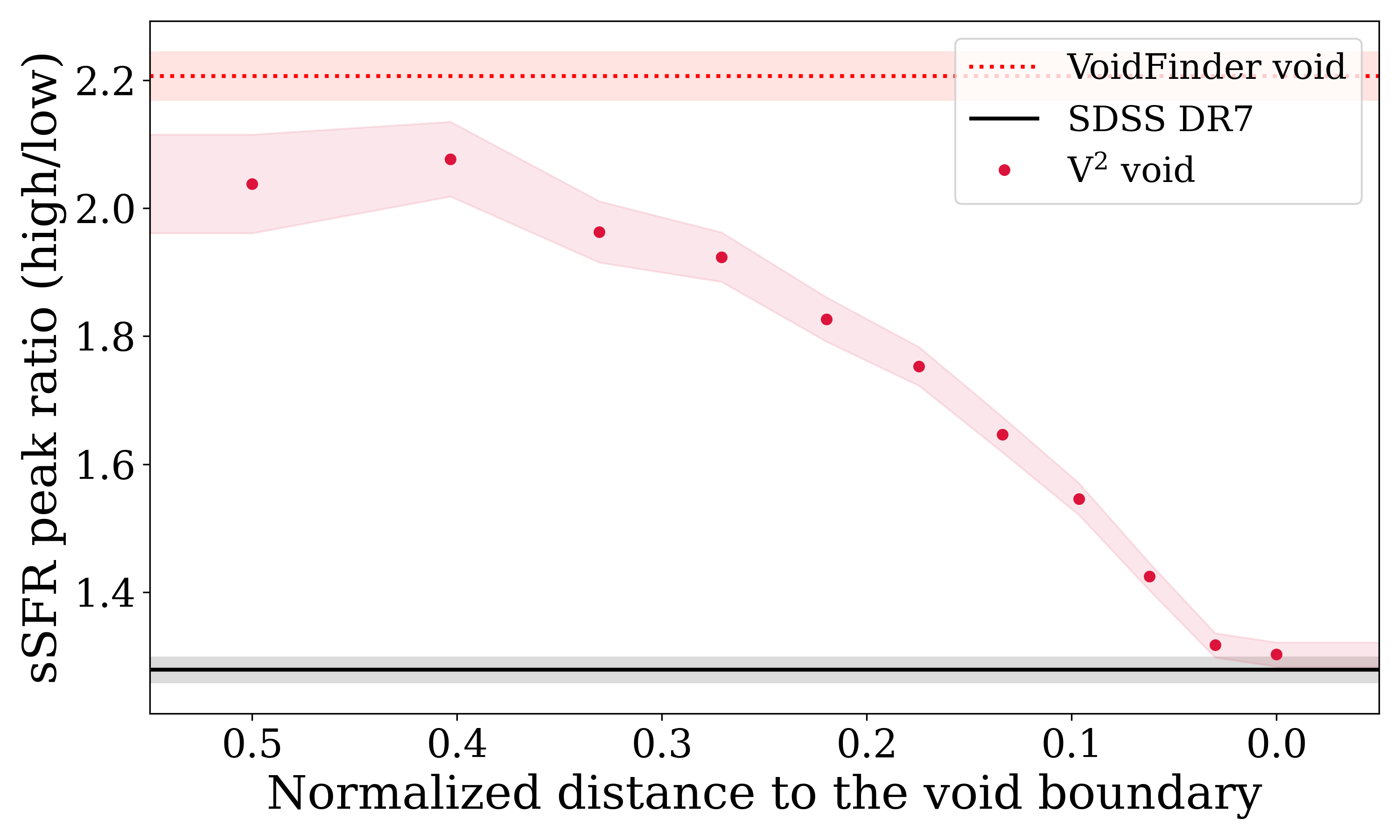}
  \caption{The relationship between the distribution of sSFR of void galaxies, 
  as classified by \VV, and the galaxies' depths within the voids in which they 
  reside (whether they live close to the edge of the void or deep inside it).  
  The deepest galaxies within each void have a normalized distance of 1, while a 
  distance closer to 0 indicates galaxies that are close to the edge.  Galaxies 
  further from the edges of the \VV voids have sSFR distributions which more 
  closely match that of \VF void galaxies; as galaxies closer to the \VV void 
  boundaries are included in the sample, the sSFR distribution shifts to more 
  closely resemble the SDSS~DR7 galaxy distribution.} 
  \label{fig:sfr_depth}
\end{figure}

If the void samples of \VV are contaminated by wall galaxies located at the 
edges of its voids, the expected void galaxy properties can be recovered by 
focusing on the galaxies that are closer to the centers of its voids.  Studying 
the galaxy properties as a function of void depth, we can determine the void 
depth within which \VV void galaxies have similar trends to \VF void galaxies.  
This is the observational complement to one of the mitigation techniques 
discussed in \cite{Veyrat23}, who tested this idea on the dynamical 
classification of simulated dark matter particles.

We calculate a galaxy's distance to the void's edge to determine how deep a 
galaxy is within a \VV void.  This is equivalent to the distance from the 
position of the galaxy to the edge of the void, defined by the Voronoi-cell 
surfaces that make up the boundary between a cell inside the void and a cell 
outside of it.  We normalize the depth of each galaxy by the largest depth 
calculated for that void, so that we can compare voids of different sizes.  
Therefore, the deepest galaxy within each void has a depth of 1, and a galaxy 
close to the edge will have a depth close to 0.

To test this hypothesis, we fit subsets of the \VV void galaxies with a skew 
normal (or sum of two skew normals if the distribution is bimodal) as a function 
of void depth for each galaxy property studied in 
Section~\ref{sec:influence_results}.  For a given void depth, the corresponding 
subsample of void galaxies consists of all galaxies with depths greater than the 
given depth threshold.  This allows us to study the behavior of the \VV void 
galaxy population if the voids were limited to some inner fraction of the 
current total void size.  The results of this analysis for the sSFR is shown in 
Figure~\ref{fig:sfr_depth}.  For comparison, we also include the ratio of the 
sSFR peak amplitudes for \VF void galaxies (in red) and the ratio of the sSFR 
peak amplitudes for all galaxies in the overall NSA catalog (in black).

For unimodal galaxy properties (stellar mass and luminosity), we find a steady 
shift from smaller masses / fainter luminosities to larger masses / brighter 
luminosities as the \VV void galaxies include more of those closer to the void 
boundaries.  For properties with bimodal distributions (color and (s)SFR), the 
fraction of blue galaxies / galaxies with higher (s)SFR decreases while the 
fraction of red galaxies / galaxies with lower (s)SFR increases as we include 
more \VV void galaxies closer to their void boundaries.  
Figure~\ref{fig:sfr_depth} shows the ratio of the peak of the mode at higher 
sSFR to the mode at lower sSFR as a function of minimum depth included in the 
subsample.

For all properties, we find that the \VV void galaxy distribution matches that 
of \VF void galaxies when only \VV void galaxies in the inner $\sim$20\% of the 
volume of the \VV voids are included in the void sample.  Therefore, the 
galaxies sitting close to the boundaries of \VV voids are more likely to be wall 
galaxies and not actual void galaxies.  This matches the results of 
\cite{Habouzit20}, who apply \texttt{VIDE} to a cosmological hydrodynamical 
simulation and reproduce the expected void galaxy property results when 
including only those galaxies closest to the centers of the \texttt{VIDE} voids.  
This also matches the simulation results of \cite{Veyrat23}, who compare the 
distribution of dark matter particles with a given crossing number as a function 
of void depth for both \VF and \VV.  \cite{Nadathur14} also discusses the 
tendency for the watershed algorithm to include high-density walls and filaments 
within void regions.

Therefore, our analysis of galaxy properties as a function of depth within \VV 
voids is consistent with our hypothesis and previous studies that the void 
sample of \VV is contaminated by wall galaxies, a result of the extension of \VV 
voids into the walls.  This is visible on the right in 
Figure~\ref{fig:DR7_slice}, where clustered void galaxies (in red) are located 
at the void edges.  This phenomenon is also visible in our mock 2D galaxy 
catalog (see the middle and right panels of Figure~\ref{fig:vv_2D}), where 
multiple cells outside the artificial voids are joined to the void region.  This 
``leaking'' of the void regions into the walls is a consequence of how the 
watershed algorithm within \VV (and other \texttt{ZOBOV}-like void-finders) 
defines void boundaries based on density maxima.  To use \VV voids to study 
galaxy properties, one must therefore implement a depth limit to exclude wall 
galaxies from the void galaxy population.

\section{Conclusions}

We examine the impact of the choice of void-finding algorithm used to classify 
the large-scale structure of galaxies by comparing properties of galaxies 
determined to be void galaxies using the \VF and \VV algorithms.  Using the \VF 
void catalog for SDSS~DR7, we find that the distributions of void galaxy 
properties are consistent with the expected physical properties of galaxies 
existing in voids: smaller masses, fainter luminosities, bluer colors, and 
higher (specific) star formation rates.  When using \VV, however, we find very 
little difference between the properties of void galaxies and the properties of 
galaxies in denser regions.

Upon further investigation, we find that \VV void regions include a significant 
fraction of the walls and filaments.  This is a result of the watershed 
algorithm extending void regions to the galaxy density maxima, which are located 
in the centers of the denser regions in the galaxy distribution.  While \VF is 
better suited for studying void galaxy properties because it is not subject to 
this phenomenon, it is possible to recover the void galaxy population with \VV 
voids if voids are limited to the inner $\sim$20\% of their volume.

\section{Acknowledgements}

The authors thank Stephen W. O'Neill, Jr. for his help and expertise with \VAST, 
as well as Michael Vogeley, Regina Demina, and Tom Ferbel for their feedback on 
this manuscript. D. Veyrat and S. BenZvi acknowledge support from the U.S. 
Department of Energy Office of High Energy Physics under the Award Number 
DE-SC0008475.

Funding for the SDSS and SDSS-II has been provided by the Alfred P. Sloan 
Foundation, the Participating Institutions, the National Science Foundation, the 
U.S. Department of Energy, the National Aeronautics and Space Administration, 
the Japanese Monbukagakusho, the Max Planck Society, and the Higher Education 
Funding Council for England.  The SDSS web site is \url{http://www.sdss.org/}.

The SDSS is managed by the Astrophysical Consortium for the Participating 
Institutions.  The Participating Institutions are the American Museum of Natural 
History, Astrophysical Institute Potsdam, University of Basel, University of 
Cambridge, Case Western Reserve University, University of Chicago, Drexel 
University, Fermilab, the Institute for Advanced Study, the Japan Participation 
Group, Johns Hopkins University, the Joint Institute for Nuclear Astrophysics, 
the Kavli Institute for Particle Astrophysics and Cosmology, the Korean 
Scientist Group, the Chinese Academy of Sciences (LAMOST), Los Alamos National 
Laboratory,  the Max-Planck-Institute for Astronomy (MPIA), the 
Max-Planck-Institute for Astrophysics (MPA), New Mexico State University, Ohio 
State University, University of Pittsburgh, University of Portsmouth, Princeton 
University, the United States Naval Observatory, and the University of 
Washington.


\appendix

\section{VAST examples}

\begin{figure}
  \includegraphics[width=0.5\textwidth]{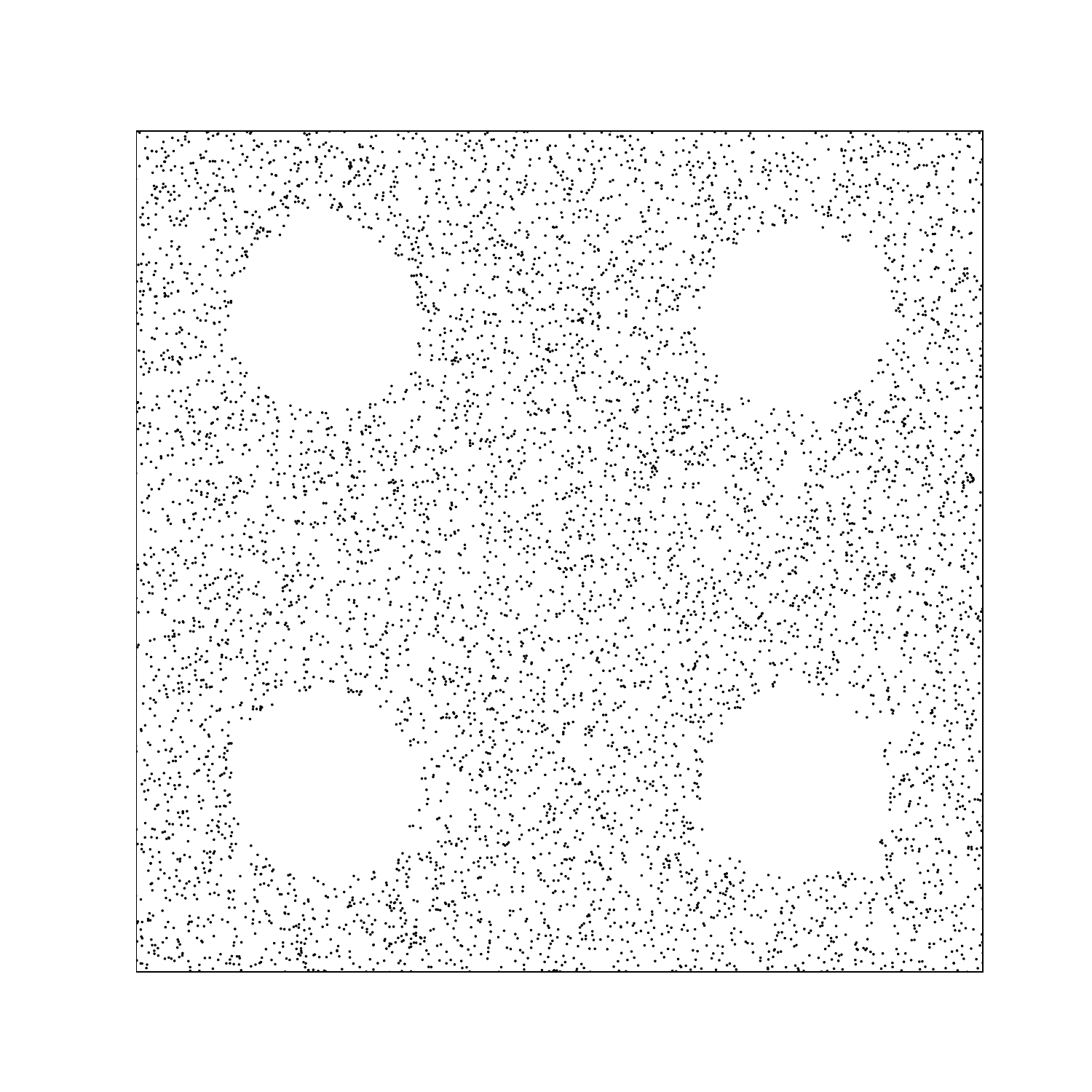}
  \caption{A sample random 2D galaxy catalog with circles of radius $10\bar{r}$ 
  removed in each quadrant to simulate simple voids, where $\bar{r}$ is the mean 
  inter-particle separation.}
  \label{fig:empty}
\end{figure}

To visualize how the two algorithms included in \VAST identify voids, we apply 
each algorithm to a two-dimensional (2D) mock catalog of randomly distributed 
galaxies.  The artificial voids are formed by emptying four circular holes in 
the random galaxy distribution, as shown in Figure~\ref{fig:empty}.

\subsection{\VF example}\label{sec:VF_appendix}

\begin{figure*}
  \centering
  \includegraphics[width=0.32\textwidth]{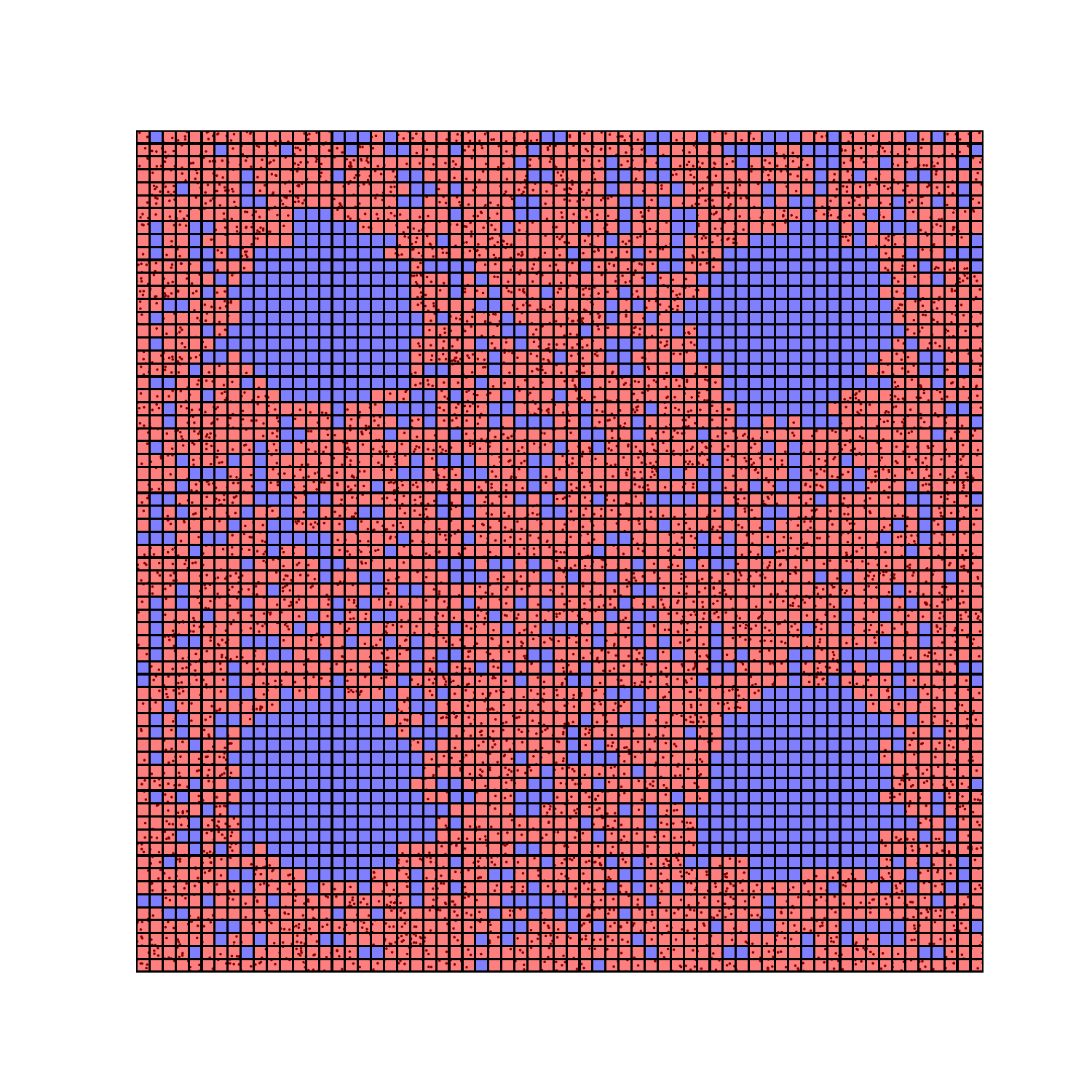}
  \includegraphics[width=0.32\textwidth]{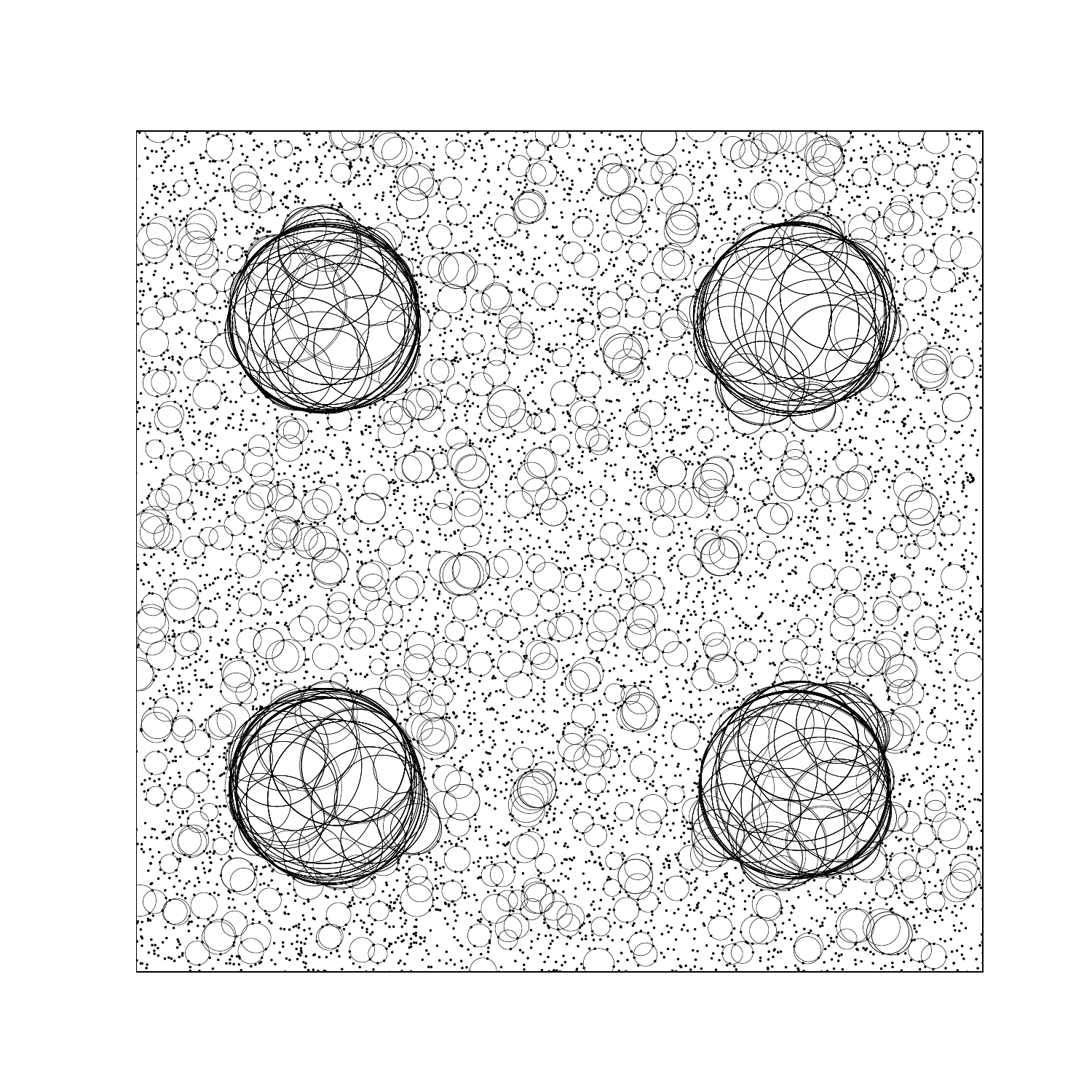}
  \includegraphics[width=0.32\textwidth]{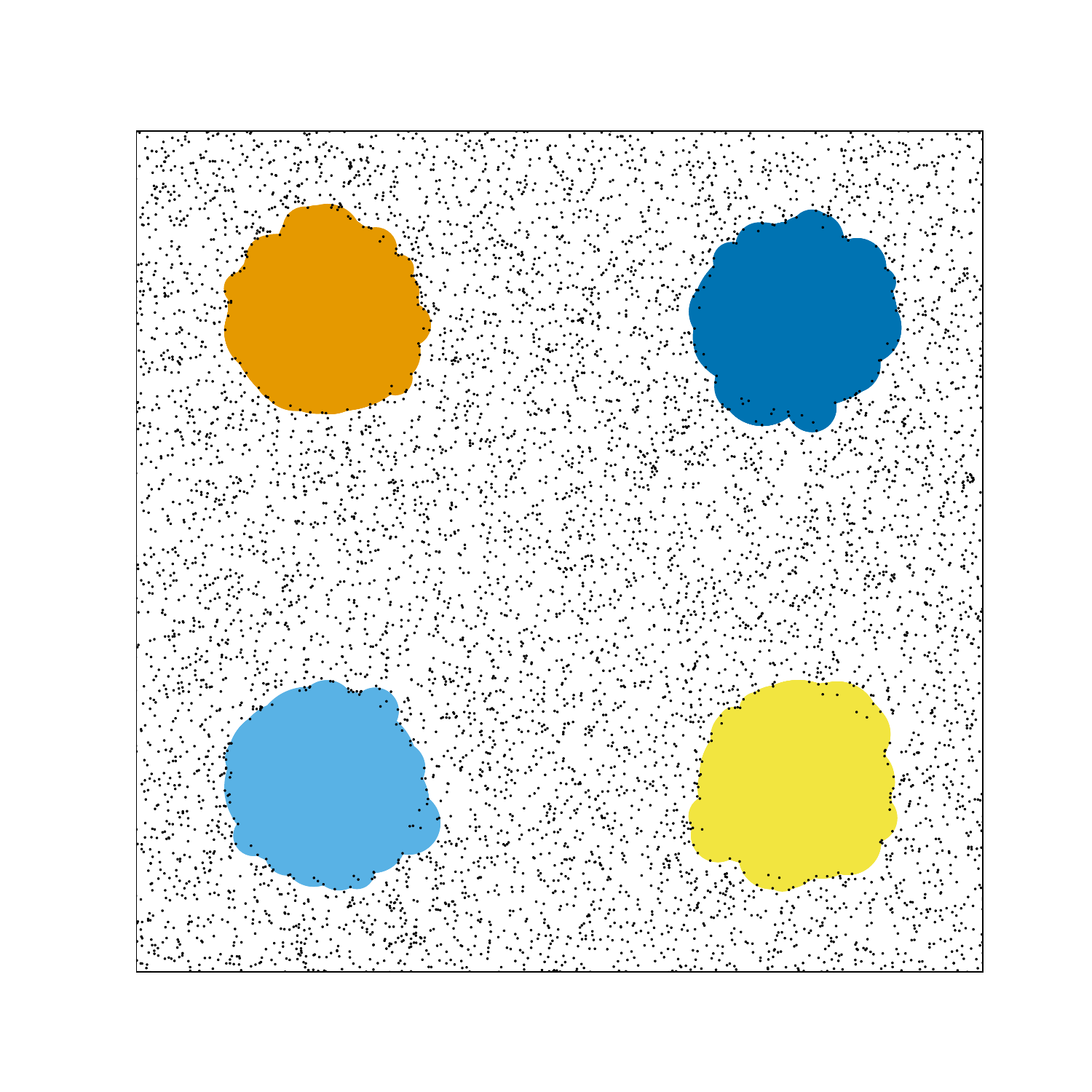}
  \caption{Steps in the \VF algorithm applied to a 2D mock galaxy catalog.  
  \emph{Left:} Non-field (wall) galaxies placed on a grid.  Any grid cell 
  containing at least one galaxy is shaded red; all empty grid cells are blue.  
  \emph{Center:} Circles grown from the center of each empty grid cell until 
  bounded by at least three galaxies.  
  \emph{Right:} Voids formed through the union of the grown spheres.}
  \label{fig:vf_2D}
\end{figure*}

After removing isolated (field) galaxies, the remaining (wall) galaxies are 
placed on a 5\hMpc grid.  An example of this grid is shown on the left in 
Figure~\ref{fig:vf_2D} for the 2D mock galaxies, where empty cells are colored blue.

Next, a sphere (or hole) is expanded from each empty cell in the grid.  Each 
sphere, centered on the cell, is created with radius equal to the distance to 
the nearest wall galaxy. The sphere's center then shifts away from that galaxy 
as its radius increases until its surface encounters a second galaxy.  At this 
point, the sphere expands from the midpoint between these two galaxies until a 
third galaxy is encountered. Finally, the sphere grows perpendicularly away from 
the plane formed by these three galaxies until it reaches a fourth galaxy.  The 
center panel of Figure~\ref{fig:vf_2D} shows the result of this step for the 2D 
catalog of mock galaxies.

To extract unique voids from this list of holes, the holes are sorted by radius, 
and the largest hole is identified as a maximal sphere --- the largest sphere 
that can fit in a given void region. Then, for each subsequent hole, if it does 
not overlap any already-identified maximal spheres by more than 10\% of its 
volume, it is also identified as a maximal sphere.  This process continues for 
all holes with radii larger than $10\bar{r}$, where $\bar{r}$ is the mean 
particle separation in the simulated random sample.

Finally, the volume of the void is refined by merging each maximal sphere with 
the remaining non-maximal spheres. When a hole overlaps only one maximal sphere 
by more than 50\% of its volume, it is merged into that void.  Holes that 
overlap multiple maximal spheres by more than 50\% are discarded.  Each void is 
therefore composed of one maximal sphere and some number of smaller holes.  The 
right panel in Figure~\ref{fig:vf_2D} shows the final 2D catalog of voids found 
using \VF.

\subsection{\VV example}\label{sec:V2_appendix}

\begin{figure*}
  \centering
  \includegraphics[width=0.32\textwidth]{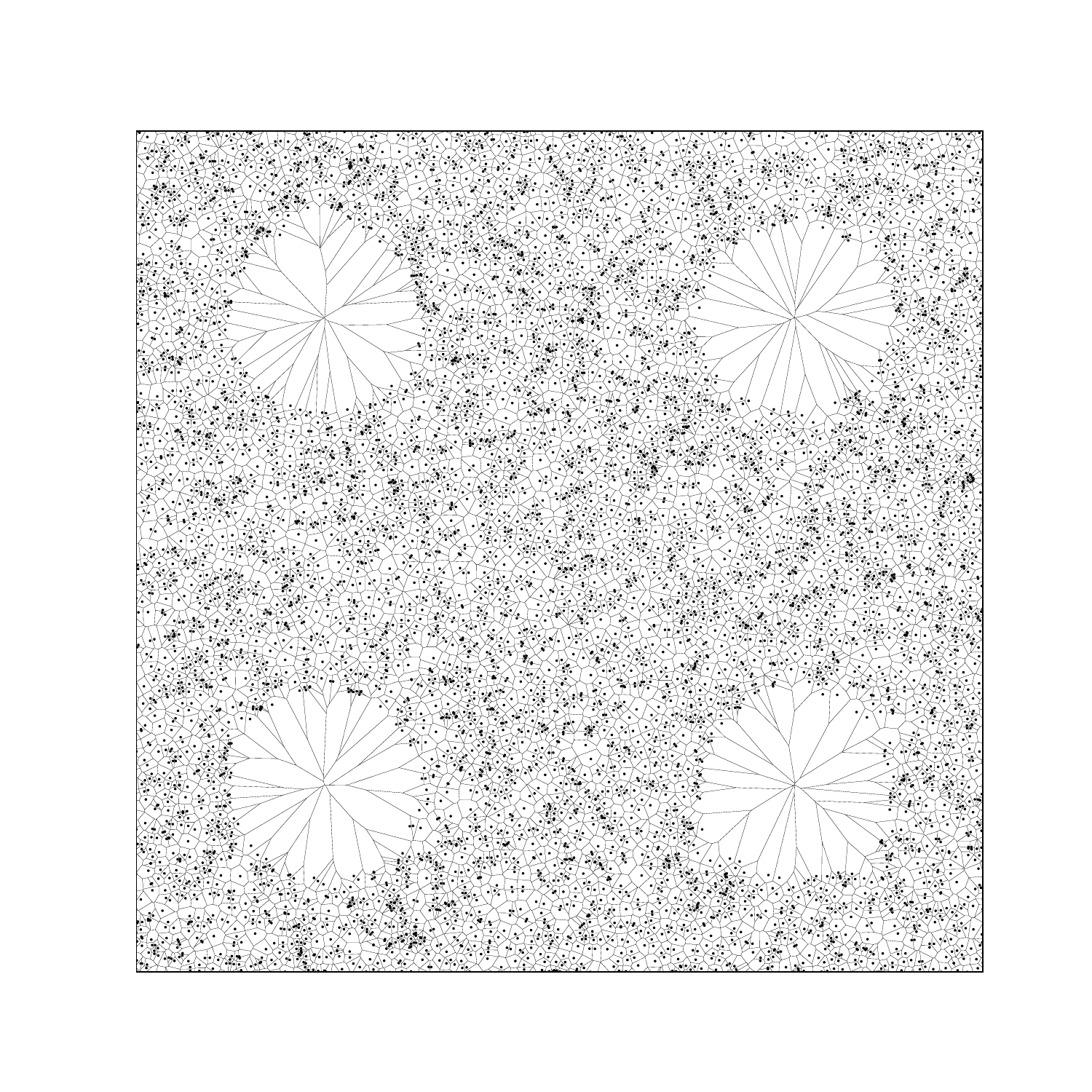}
  \includegraphics[width=0.32\textwidth]{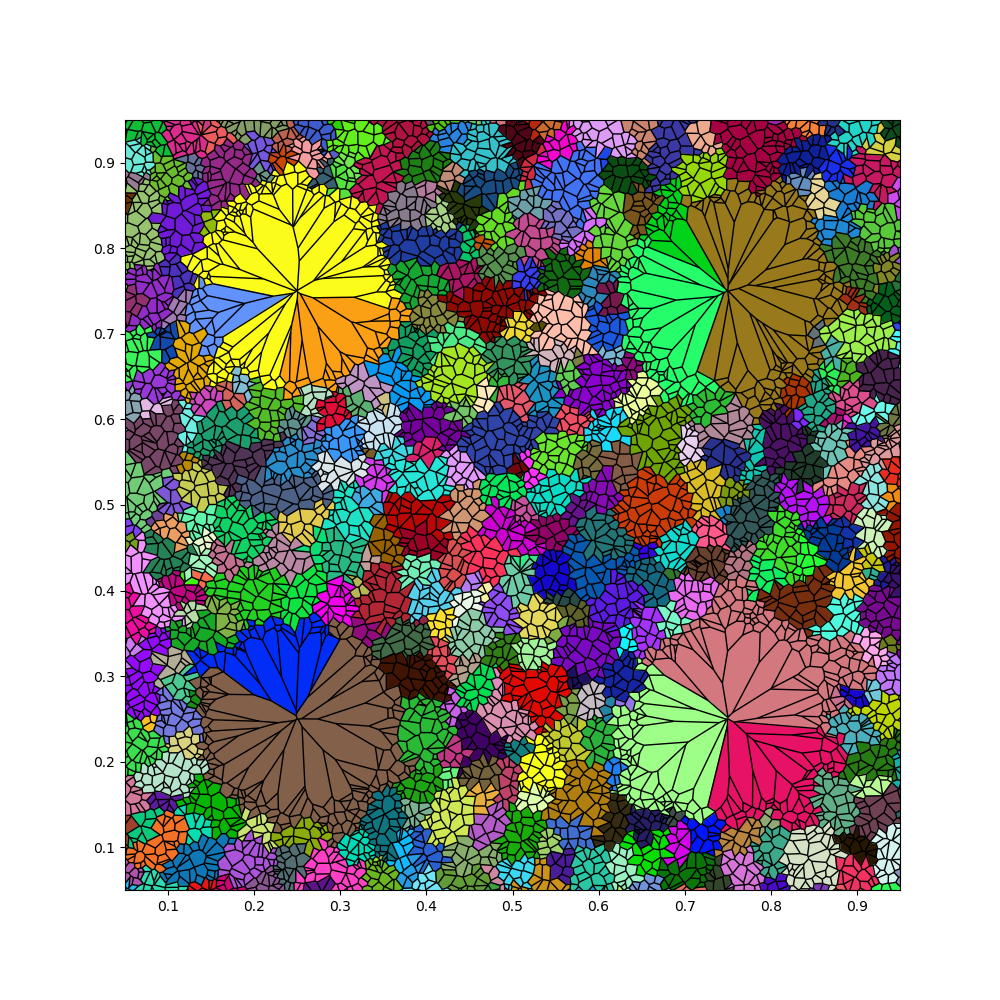}
  \includegraphics[width=0.32\textwidth]{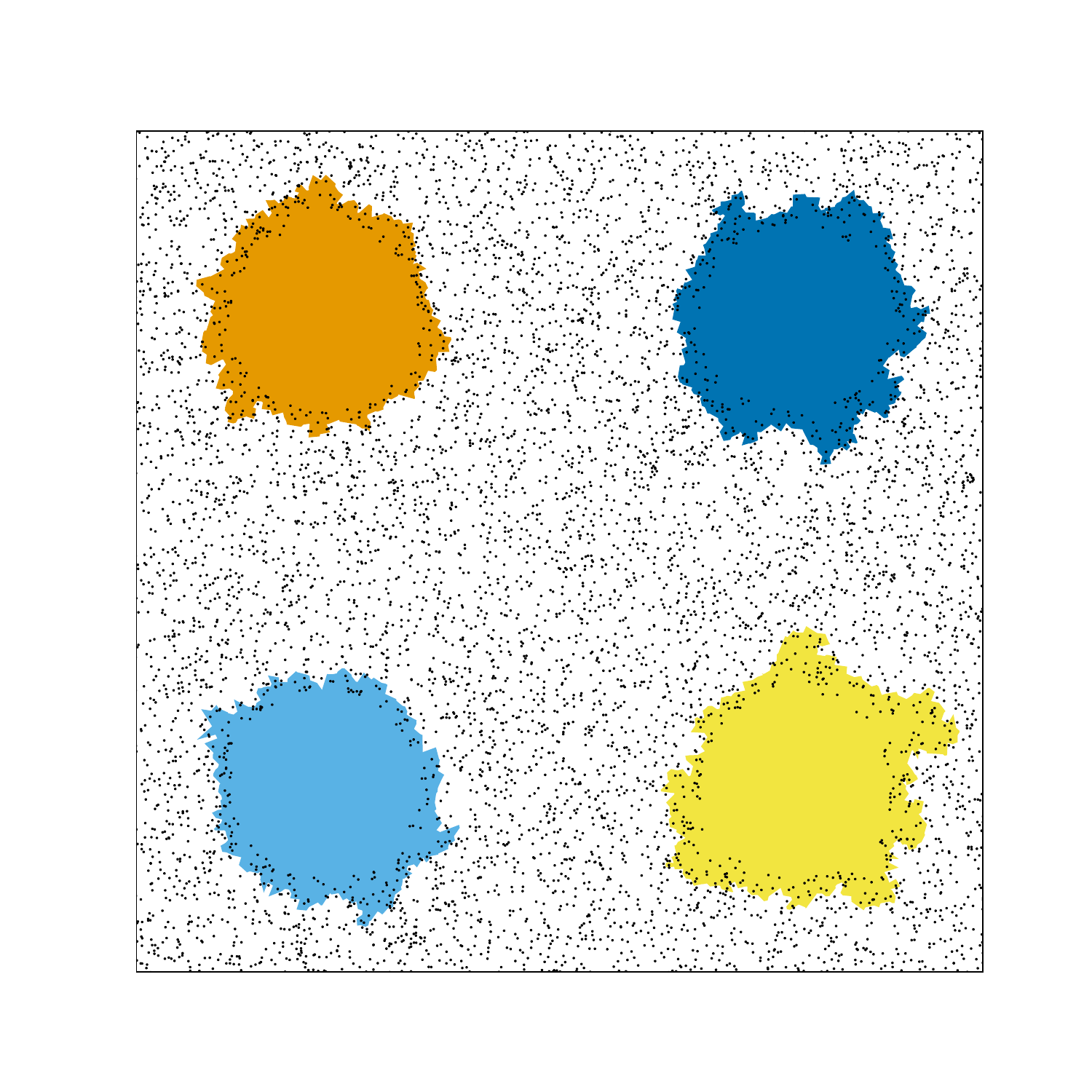}
  \caption{Steps in the \VV algorithm applied to a 2D mock galaxy catalog.  
  \emph{Left:} The Voronoi tessellation of the catalog.  
  \emph{Center:} Zones formed using a watershed method.  
  \emph{Right:} Final voids identified after merging zones and pruning voids.}
  \label{fig:vv_2D}
\end{figure*}

\VV can be divided into four steps:
\begin{enumerate}\setlength{\itemsep}{-0.25em}

  \item{\bf Tessellation:} A Voronoi tessellation of the input catalog is 
  produced. The tessellation of the 2D mock galaxies is shown on the left in 
  Figure~\ref{fig:vv_2D}.  The inverse of each cell's volume is an estimate of 
  that galaxy's local density.
    
  \item{\bf Zone formation:} Voronoi cells are combined into groups or ``zones'' 
  using watershed segmentation.  Each cell is put into the same zone as its 
  least-dense neighbor, and cells less dense than any neighboring cell (local 
  density minima) are identified as their zone’s central cell.  The center panel 
  of Figure~\ref{fig:vv_2D} shows the zones formed using the 2D mock sample.
    
  \item{\bf Zone merging:} Zones are merged to form voids by first identifying 
  the least-dense pair of adjacent cells between two zones.  For a given maximum 
  ``linking density,'' the set of all zones is partitioned into a subset of 
  voids linked together by zones of equal or lower density.  The algorithm then 
  loops over all linking densities to produce a list of voids.
    
  \item{\bf Void pruning:} Zone linking produces a hierarchy of voids ranging 
  from individual zones to a single void encompassing the entire survey volume.  
  A pruning step similar to that implemented in \texttt{VIDE} \citep{Sutter15} 
  is therefore introduced to identify the final list of voids.  The final result 
  for the 2D sample with this pruning is shown on the right in 
  Figure~\ref{fig:vv_2D}.
  
\end{enumerate}

\subsection{Calculating the Bayes factor}\label{sec:BF_appendix}

Due to our large sample size, non-Bayesian significance tests such as a 
two-sample Kolmogorov–Smirnov (K-S) test yield extremely low p-values that 
inaccurately quantify the extent of similarity of our distributions.  Since we 
know that our data is distributed like one or more Gaussian or skew normal 
distributions (a mixture model), we can bin and fit the data to the model 
assuming that the samples come from either the same parent distribution or two 
different parent distributions. Then, we can compare these two models using the 
Bayes Factor.

We have two data sets with binned counts $\mathbf{m} = {m_i}$ and 
$\mathbf{n} = {n_i}$.  For the single-parent model, we assume that both data 
sets are created by the same skew normal mixture model, such that the mean count 
in bin $i$ for data set 1 is
\begin{equation}\label{eq:meancount}
  \lambda_{1,i} = \left[\alpha\ \mathrm{SN}(\mu_\alpha,\sigma_\alpha,\xi_\alpha) + \beta\ \mathrm{SN}(\mu_\beta,\sigma_\beta,\xi_\beta)\right]_i.
\end{equation}
To account for the different sizes in the data sets, we can add a scale factor 
$s$ to the fit of the second data set:
\begin{equation}\label{eq:scaledmean}
  \lambda_{2,i} = \left[s\left(\alpha\ \mathrm{SN}(\mu_\alpha,\sigma_\alpha,\xi_\alpha) + \beta\ \mathrm{SN}(\mu_\beta,\sigma_\beta,\xi_\beta)\right)\right]_i = s\lambda_{1,i}.
\end{equation}
Thus the model includes 9 parameters 
$\theta=\left\{s,\alpha,\mu_\alpha,\sigma_\alpha,\xi_\alpha,\beta,\mu_\beta,\sigma_\beta,\xi_\beta\right\}$.  
We can then maximize the joint Poisson likelihood
\begin{equation}\label{eq:maxlike}
  p(\mathbf{m},\mathbf{n}|\lambda_1,\lambda_2,\mathcal{M}_1)
  = \prod_{i=1}^N \frac{\lambda_{1,i}^{m_i}e^{-\lambda_{1,i}}}{m_i!}
  \frac{\lambda_{2,i}^{n_i}e^{-\lambda_{2,i}}}{n_i!},
\end{equation}
or minimize the negative log likelihood
\begin{equation}\label{eq:loglike}
  -\ln{\mathcal{L}} = -\sum_{i=1}^N \left(m_i\lambda_{1,i}-\lambda_{1,i}-\ln{m_i!} + 
  n_i\lambda_{2,i}-\lambda_{2,i}-\ln{n_i!}\right)
\end{equation}
to obtain the best-fit parameters.

If on the other hand we assume the binned counts $\mathbf{m} = {m_i}$ and 
$\mathbf{n} = {n_i}$ are created by separate skew normal mixture models with 
mean count
\begin{equation}\label{eq:meanM2_i}
  \lambda_{1,i} = \left[\alpha\ \mathrm{SN}(\mu_\alpha,\sigma_\alpha,\xi_\alpha) + \beta\ \mathrm{SN}(\mu_\beta,\sigma_\beta,\xi_\beta)\right]_i
\end{equation}
in set 1 and mean count
\begin{equation}\label{eq:meanM2_ii}
  \lambda_{2,i} = \left[\gamma\ \mathrm{SN}(\mu_\gamma,\sigma_\gamma,\xi_\gamma) + \delta\ \mathrm{SN}(\mu_\delta,\sigma_\delta,\xi_\delta)\right]_i
\end{equation}
in set 2, then the model includes 16 free parameters
\[
\theta=\left\{\alpha,\mu_\alpha,\sigma_\alpha,\xi_\alpha,\beta,\mu_\beta,\sigma_\beta,\xi_\beta,\gamma,\mu_\gamma,\sigma_\gamma,\xi_\gamma,\delta,\mu_\delta,\sigma_\delta,\xi_\delta\right\}.
\]
We can still use the likelihood in Eqn.~\ref{eq:maxlike} with mean counts from 
Eqns.~\ref{eq:meanM2_i} and \ref{eq:meanM2_ii}.

To evaluate the support of the Bayes factor for the one or two-parent 
hypothesis, we use the thresholds suggested in \cite{Jeffreys61} and 
\cite{Kass95}.

\bibliographystyle{aasjournal}
\bibliography{Zaid0324_sources}

\end{document}

%% file: t1.tex
\floattable
\begin{deluxetable*}{ccCCccc}
  \tablewidth{0pt}
  \tablecolumns{7}
  \tablecaption{Galaxy property distribution statistics\label{tab:base_model}}
  \tablehead{\colhead{Algorithm} & \colhead{Environment} & \colhead{Average} & \colhead{Median} & \colhead{Average shift} & \colhead{Median shift} & \colhead{$\log_{10}{(B_{12})}$}}
    
  \startdata
    \cutinhead{Stellar mass [$\log (M_*/M_\odot)$]}
    \multirow{2}{*}{\VF} & Void & 9.786 \pm 0.002 & 9.910 & \multirow{2}{*}{$0.205 \pm 0.002\text{ (}2.1 \pm 0.2$\%)} & \multirow{2}{*}{$0.164\text{ (}1.6$\%)} & \multirow{2}{*}{$-1708$}\\
                         & Wall & 9.992 \pm 0.001 & 10.074 & & & \\
    \hline
    \multirow{2}{*}{\VV} & Void & 9.927 \pm 0.001 & 10.025 & \multirow{2}{*}{$0.038 \pm 0.002\text{ (}0.4 \pm 0.2$\%)} & \multirow{2}{*}{$0.043\text{ (}0.4$\%)} & \multirow{2}{*}{$-203$}\\
                         & Wall & 9.965 \pm 0.002 & 10.068 & & & \\
    \cutinhead{Absolute magnitude, $M_r$}
    \multirow{2}{*}{\VF} & Void & -19.619 \pm 0.004 & -19.837 & \multirow{2}{*}{$-0.359 \pm 0.005\text{ (}1.8 \pm 0.5$\%)} & \multirow{2}{*}{$-0.263\text{ (}1.3$\%)} & \multirow{2}{*}{$-1248$}\\
                         & Wall & -19.978 \pm 0.002 & -20.100 & & & \\
    \hline
    \multirow{2}{*}{\VV} & Void & -19.859 \pm 0.002 & -20.013 & \multirow{2}{*}{$-0.075 \pm 0.005\text{ (}0.4 \pm 0.5$\%)} & \multirow{2}{*}{$-0.088\text{ (}0.4$\%)} & \multirow{2}{*}{$-308$}\\
                        & Wall & -19.934 \pm 0.004 & -20.101 & & & \\
    \cutinhead{$u-r$}
    \multirow{2}{*}{\VF} & Void & 1.763 \pm 0.002 & 1.730 & \multirow{2}{*}{$0.173 \pm 0.002\text{ (}9.0 \pm 0.2$\%)} & \multirow{2}{*}{$0.259\text{ (}13$\%)} & \multirow{2}{*}{$-1592$}\\
                         & Wall & 1.937 \pm 0.001 & 1.989 & & & \\
    \hline
    \multirow{2}{*}{\VV} & Void & 2.102 \pm 0.001 & 2.127 & \multirow{2}{*}{$0.027 \pm 0.002\text{ (}1.3 \pm 0.2$\%)} & \multirow{2}{*}{$0.039\text{ (}1.8$\%)} & \multirow{2}{*}{$-28.8$}\\
                         & Wall & 2.130 \pm 0.002 & 2.166 & & & \\
    \cutinhead{$g-r$}
    \multirow{2}{*}{\VF} & Void & 0.5614 \pm 0.0007 & 0.5729 & \multirow{2}{*}{$0.0540 \pm 0.0008\text{ (}8.77 \pm  0.08$\%)} & \multirow{2}{*}{$0.0855\text{ (}13.0$\%)} & \multirow{2}{*}{$-1704$}\\
                         & Wall & 0.6154 \pm 0.0004 & 0.6584 & & & \\
    \hline
    \multirow{2}{*}{\VV} & Void & 0.6836 \pm 0.0005 & 0.7201 & \multirow{2}{*}{$0.0128 \pm 0.0009\text{ (}1.84 \pm 0.09$\%)} & \multirow{2}{*}{$0.0172\text{ (}2.33$\%)} & \multirow{2}{*}{$-57.9$}\\
                         & Wall & 0.6963 \pm 0.0007 & 0.7373 & & & \\
    \cutinhead{SFR [$\log (M_\odot/\text{yr})$]}
    \multirow{2}{*}{\VF} & Void & -0.303 \pm 0.002 & -0.216 & \multirow{2}{*}{$-0.097 \pm 0.003\text{ (}24.2 \pm 0.3$\%)} & \multirow{2}{*}{$-0.175\text{ (}44.8$\%)} & \multirow{2}{*}{$-416$}\\
                         & Wall & -0.400 \pm 0.001 & -0.391 & & & \\
    \hline
    \multirow{2}{*}{\VV} & Void & -0.385 \pm 0.001 & -0.355 & \multirow{2}{*}{$0.008 \pm 0.003\text{ (}2.1 \pm 0.3$\%)} & \multirow{2}{*}{$0.020\text{ (}5.9$\%)} & \multirow{2}{*}{$-66.5$}\\
                         & Wall & -0.377 \pm 0.002 & -0.336 & & & \\
    \cutinhead{sSFR [$\log (\text{yr}^{-1})$]}
    \multirow{2}{*}{\VF} & Void & -10.489 \pm 0.003 & -10.232 & \multirow{2}{*}{$-0.294 \pm 0.003\text{ (}2.7 \pm 0.3$\%)} & \multirow{2}{*}{$-0.445\text{ (}4.2$\%)} & \multirow{2}{*}{$-1471$}\\
                         & Wall & -10.783 \pm 0.002 & -10.677 & & & \\
    \hline
    \multirow{2}{*}{\VV} & Void & -10.705 \pm 0.002 & -10.521 & \multirow{2}{*}{$-0.044 \pm 0.003\text{ (}0.4 \pm 0.3$\%)} & \multirow{2}{*}{$-0.094\text{ (}0.9$\%)} & \multirow{2}{*}{$-18.9$}\\
                        & Wall & -10.750 \pm 0.003 & -10.615 & & & 
  \enddata
    
  \tablecomments{Galaxy property distribution summary for void and wall 
  environments according to \VF or \VV.  The shifts are calculated as $($wall 
  $-$ void$)$, and the percentages are equal to 
  $(\text{wall} - \text{void})/\text{wall}$, with a negative sign indicating 
  that void galaxies have larger values than the wall galaxies. The logarithm of 
  the Bayes factor is included; all Bayes factors are significantly less than 
  $\log(1/100) = -2$, indicating definitive evidence for the two-parent model 
  (Appendix~\ref{sec:BF_appendix}).}
\end{deluxetable*}